%% file: Main.tex
\documentclass[twocolumn,twocolappendix,trackchanges]{aastex702}
\usepackage[version=4]{mhchem}
\newcommand{\rah}{\ensuremath{^{\rm h}}}
\newcommand{\ram}{\ensuremath{^{\rm m}}}

\begin{document}
\title{The ALMA-QUARKS survey: A Keplerian Circumbinary Disk Candidate in the High-mass Star-forming Region IRAS 08303--4303}

\input{0_author}


\begin{abstract}
We present the ALMA Band~6 continuum and molecular line observations of IRAS 08303--4303, a high-mass star-forming region, with an angular resolution of $\sim$0\farcs3 ($\sim$700~au at the distance of 2.35~kpc). 
Six dense condensations are identified in the 1.3~mm continuum, and at least four of them are associated with molecular outflows. 
The estimated outflow parameters, particularly mass outflow rates $\geq10^{-4}~M_{\sun}~\mathrm{yr}^{-1}$, suggest that the driving sources are high-mass young stellar objects. 
Two of the condensations are forming a wide protobinary system with a separation of $\sim$900~au. 
They reside within a single rotating structure, identified as a circumbinary disk which has a mass of about 2~$M_{\sun}$. 
This structure is traced by multiple molecular lines and exhibits a Keplerian rotation pattern. 
Based on the Keplerian rotation assumption, the enclosed mass is derived to be between 32 and 61~$M_{\sun}$: 22--43~$M_{\sun}$ for the primary and 10--18~$M_{\sun}$ for the secondary. 
The circumbinary disk is on the whole gravitationally stable, yet some parts with Toomre $Q < 2$ indicate potential fragmentation. 
Various molecular lines reveal a streamer-like structure connecting to the circumbinary disk, suggesting transfer of material from the envelope to the protobinary system through it. 
Despite the observational rarity of such circumbinary disks around high-mass YSOs, the high binary fraction among massive stars underscores their significance. 
This source, therefore, serves as a key laboratory for understanding the formation of high-mass binary systems. 
\end{abstract}



\section{Introduction} \label{sec:intro}
\input{1_intro}

\section{Observation} \label{sec:obs}
\input{2_obs}

\section{Results} \label{sec:result}

\subsection{Dense Continuum Condensations} \label{subsec:result-cont}
\input{3_1_cont}

\subsection{Molecular Outflows} \label{subsec:result-outflow}
\input{3_2_outflow}

\subsection{Circumbinary Disk Candidate around the Protobinary System} \label{subsec:result-disk}
\input{3_3_diskC1C2}

\subsection{Disk Candidate Associated with the Other Condensations} \label{subsec:result-diskSW2}
\input{3_4_diskSW2}

\section{Discussion} \label{sec:discuss}

\subsection{Kinematics of the Circumbinary Disk} \label{subsec:discuss-disk}
\input{4_1_kinematic}

\subsection{Comparison with Previous Studies} \label{subsec:discuss-compare}
\input{4_2_previous}

\subsection{Molecular Streamer} \label{subsec:discuss-streamer}
\input{4_3_streamer}

\subsection{Nature of the IRAS 08303--4303} \label{subsec:discuss-nature}
\input{4_4_nature}

\section{Summary} \label{sec:sum}
\input{5_Summary}

\begin{acknowledgments}
\input{6_acknowledgments}
\end{acknowledgments}

\facilities{ALMA}
\software{CASA \citep{2022PASP..134k4501C}, 
Keplerfit \citep{2019A&A...629A..10B, 2023ascl.soft08012B}
}


\restartappendixnumbering
\appendix

\section{Temperature calculation} \label{sec:appendixA}
\input{7_Temprot}

\section{\texorpdfstring{\ce{CH3CN}}{CH3CN} velocity dispersion}
\label{sec:appendixB}
\input{8_ch3cn}

\bibliography{ref}{}
\bibliographystyle{aasjournalv7.1}

\end{document}

%% file: 0_author.tex
\author[0009-0002-9273-7220]{Gyuho Lee}
\email[show]{space295@kasi.re.kr}
\affiliation{Korea Astronomy and Space Science Institute, 776 Daedeokdae-ro, Yuseong-gu, Daejeon 34055, Republic of Korea}
\affiliation{University of Science and Technology (UST), 217 Gajeong-ro, Yuseong-gu, Daejeon 34113, Republic of Korea}

\author[0000-0003-2412-7092]{Kee-Tae Kim}
\email[show]{ktkim@kasi.re.kr}
\affiliation{Korea Astronomy and Space Science Institute, 776 Daedeokdae-ro, Yuseong-gu, Daejeon 34055, Republic of Korea}
\affiliation{University of Science and Technology (UST), 217 Gajeong-ro, Yuseong-gu, Daejeon 34113, Republic of Korea}

\author{Tie Liu}
\affiliation{Shanghai Astronomical Observatory, Chinese Academy of Sciences, 80 Nandan Road, Shanghai 200030, P.\ R.\ China}
\email{liutie@shao.ac.cn}

\author{Xunchuan Liu}
\affiliation{Shanghai Astronomical Observatory, Chinese Academy of Sciences, 80 Nandan Road, Shanghai 200030, P.\ R.\ China}
\affiliation{Leiden Observatory, Leiden University, P.O.\ Box 9513, 2300RA Leiden, The Netherlands}
\email{liuxunchuan001@gmail.com}

\author{Sheng-Yuan Liu}
\affiliation{Institute of Astronomy and Astrophysics, Academia Sinica, No.1, Sec. 4, Roosevelt Rd, Taipei 106216, Taiwan (R.O.C.)}
\email{syliu@asiaa.sinica.edu.tw}

\author{Lei Zhu}
\affiliation{Chinese Academy of Sciences South America Center for Astronomy, National Astronomical Observatories, CAS, Beijing 100101, P.\ R.\ China}
\email{lzhu@das.uchile.cl}

\author[0000-0003-1649-7958]{Guido Garay}
\affiliation{Chinese Academy of Sciences South America Center for Astronomy, National Astronomical Observatories, CAS, Beijing 100101, P.\ R.\ China}
\affiliation{Departamento de Astronom\'{i}a, Universidad de Chile, Camino el Observatorio 1515, Las Condes, Santiago, Chile}
\email{guido@das.uchile.cl}

\author{Jihyun Kang}
\affiliation{Korea Astronomy and Space Science Institute, 776 Daedeokdae-ro, Yuseong-gu, Daejeon 34055, Republic of Korea}
\email{jkang@kasi.re.kr}

\author{Hyeong-Sik Yun}
\affiliation{Department of Physics and Astronomy, Seoul National University, 1 Gwanak-ro, Gwanak-gu, Seoul 08826, Republic of Korea}
\affiliation{Korea Astronomy and Space Science Institute, 776 Daedeokdae-ro, Yuseong-gu, Daejeon 34055, Republic of Korea}
\email{yun.hyeongsik@snu.ac.kr}

\author{Hong-Li Liu}
\affiliation{School of physics and astronomy, Yunnan University, Kunming, 650091, P.\ R.\ China}
\email{hongliliu2012@gmail.com}

\author[0000-0002-8697-9808]{Sami Dib}
\affiliation{Max Planck Institute for Astronomy, K\"{o}nigstuhl 17, 69117 Heidelberg, Germany}
\email{sami.dib@gmail.com}

\author[0000-0002-9875-7436]{James O.\ Chibueze}
\affiliation{UNISA Centre for Astrophysics and Space Sciences (UCASS), College of Science, Engineering and Technology, University of South Africa, Cnr Christian de Wet Rd and Pioneer Avenue, Florida Park, 1709, Roodepoort, South Africa}
\affiliation{Department of Physics and Astronomy, Faculty of Physical Sciences, University of Nigeria, Carver Building, 1 University Road, Nsukka 410001, Nigeria}
\email{james.chibueze@gmail.com}

\author{Jianjun Zhou}
\affiliation{Xinjiang Astronomical Observatory, Chinese Academy of Sciences, 150 Science 1- Street, Urumqi, Xinjiang 830011, P.\ R.\ China}
\email{zhoujj@xao.ac.cn}

\author[0000-0002-9836-0279]{Siju Zhang}
\affiliation{Departamento de Astronom\'{i}a, Universidad de Chile, Camino el Observatorio 1515, Las Condes, Santiago, Chile}
\email{sijuzhangastro@gmail.com}

\author{Di Li}
\affiliation{New Cornerstone Science Laboratory, Department of Astronomy, Tsinghua University, Beijing 100084, P.\ R.\ China}
\affiliation{National Astronomical Observatories, Chinese Academy of Sciences, Beijing 100101, P.\ R.\ China}
\email{dili@nao.cas.cn}

\author[0000-0002-8586-6721]{Pablo Garcia}
\affiliation{Chinese Academy of Sciences South America Center for Astronomy, National Astronomical Observatories, CAS, Beijing 100101, P.\ R.\ China}
\affiliation{Instituto de Astronom\'{i}a, Universidad Cat\'{o}lica del Norte, Av. Angamos 0610, Antofagasta, Chile}
\email{pablo.garcia@nao.cas.cn}

\author{Swagat R.\ Das}
\affiliation{Departamento de Astronom\'{i}a, Universidad de Chile, Camino el Observatorio 1515, Las Condes, Santiago, Chile}
\email{swagat@das.uchile.cl}

\author{Lokesh Dewangan}
\affiliation{Astronomy \& Astrophysics Division, Physical Research Laboratory, Navrangpura, Ahmedabad 380009, India}
\email{loku007@gmail.com}

\author[0000-0003-1602-6849]{Prasanta Gorai}
\affiliation{Rosseland Centre for Solar Physics, University of Oslo, P.O.\ Box 1029 Blindern, 0315 Oslo, Norway}
\affiliation{Institute of Theoretical Astrophysics, University of Oslo, P.O.\ Box 1029 Blindern, 0315 Oslo, Norway}
\email{prasanta.astro@gmail.com}

\author[0000-0002-8389-6695]{Suinan Zhang}
\affiliation{Shanghai Astronomical Observatory, Chinese Academy of Sciences, 80 Nandan Road, Shanghai 200030, P.\ R.\ China}
\affiliation{Department of Earth and Planetary Sciences, Institute of Science Tokyo, Meguro, Tokyo, 152-8551, Japan}
\affil{National Astronomical Observatory of Japan, 2-21-1 Osawa, Mitaka, Tokyo, 181-8588, Japan}
\email{suinan.zhang@gmail.com}

\author[0000-0003-4546-2623]{Aiyuan Yang}
\affiliation{National Astronomical Observatories, Chinese Academy of Sciences, Beijing 100101, P.\ R.\ China}
\affiliation{Key Laboratory of Radio Astronomy and Technology, Chinese Academy of Sciences, A20 Datun Road, Chaoyang District, Beijing, 100101, P.\ R.\ China}
\email{yangay@bao.ac.cn}

\author{Manuel Merello}
\affiliation{Departamento de Astronom\'{i}a, Universidad de Chile, Camino el Observatorio 1515, Las Condes, Santiago, Chile}
\affiliation{Centro de Astro-Ingenier\'{i}a (AIUC), Pontificia Universidad Cat\'{o}lica de Chile, Av. Vicu\~{n}a Mackena 4860, Macul, Santiago, Chile}
\email{mmerello@das.uchile.cl}

\author{Leonardo Bronfman}
\affiliation{Departamento de Astronom\'{i}a, Universidad de Chile, Camino el Observatorio 1515, Las Condes, Santiago, Chile}
\email{leo@das.uchile.cl}

\author[0000-0002-5809-4834]{Mika Juvela}
\affiliation{Department of Physics, P.O.\ Box 64, 00014, University of Helsinki, Finland}
\email{mika.juvela@helsinki.fi}

\author{Xindi Tang}
\affiliation{Xinjiang Astronomical Observatory, Chinese Academy of Sciences, 150 Science 1- Street, Urumqi, Xinjiang 830011, P.\ R.\ China}
\email{tangxindi@xao.ac.cn}

\author{Wenyu Jiao}
\affiliation{Shanghai Astronomical Observatory, Chinese Academy of Sciences, 80 Nandan Road, Shanghai 200030, P.\ R.\ China}
\email{astrojiao@gmail.com}

\author[0000-0001-5950-1932]{Fengwei Xu}
\affiliation{Max Planck Institute for Astronomy, K\"{o}nigstuhl 17, 69117 Heidelberg, Germany}
\email{fengweilookuper@gmail.com}

\author[0000-0003-4659-1742]{Zhiyuan Ren}
\affiliation{National Astronomical Observatories, Chinese Academy of Sciences, Beijing 100101, P.\ R.\ China}
\affiliation{CAS Key Laboratory of FAST, NAOC, Chinese Academy of Sciences, Beijing, P.\ R.\ China; University of Chinese Academy of Sciences, Beijing, P.\ R.\ China}
\email{renzy@nao.cas.cn}

\author{Tapas Baug}
\affiliation{S. N. Bose National Centre for Basic Sciences, Block-JD, Sector-III, Salt Lake City, Kolkata 700106, India}
\email{tapas.polo@gmail.com}

\author[0000-0001-7866-2686]{Jihye Hwang}
\affiliation{Institute for Advanced Study, Kyushu University, Japan}
\affiliation{Department of Earth and Planetary Sciences, Faculty of Science, Kyushu University, Nishi-ku, Fukuoka 819-0395, Japan}
\email{hwang.jihye.514@m.kyushu-u.ac.jp}

\author[0000-0002-7125-7685]{Patricio Sanhueza}
\affiliation{Department of Astronomy, School of Science, The University of Tokyo, 7-3-1 Hongo, Bunkyo, Tokyo 113-0033, Japan}
\email{patosanhueza@gmail.com}

\author{Yankun Zhang}
\affiliation{Shanghai Astronomical Observatory, Chinese Academy of Sciences, 80 Nandan Road, Shanghai 200030, P.\ R.\ China}
\email{zhangyankun@shao.ac.cn}

\author{Xu Zhang}
\affiliation{Shanghai Astronomical Observatory, Chinese Academy of Sciences, 80 Nandan Road, Shanghai 200030, P.\ R.\ China}
\email{zhangxu@shao.ac.cn}

\author{Chang Won Lee}
\affiliation{Korea Astronomy and Space Science Institute, 776 Daedeokdae-ro, Yuseong-gu, Daejeon 34055, Republic of Korea}
\affiliation{University of Science and Technology (UST), 217 Gajeong-ro, Yuseong-gu, Daejeon 34113, Republic of Korea}
\email{cwl@kasi.re.kr}

\author{Dongting Yang}
\affiliation{Shanghai Astronomical Observatory, Chinese Academy of Sciences, 80 Nandan Road, Shanghai 200030, P.\ R.\ China}
\email{dongting@mail.ynu.edu.cn}

\author[0000-0001-7573-0145]{Xiaofeng Mai}
\affiliation{Shanghai Astronomical Observatory, Chinese Academy of Sciences, 80 Nandan Road, Shanghai 200030, P.\ R.\ China}
\affiliation{School of Astronomy and Space Sciences, University of Chinese Academy of Sciences, No. 19A Yuquan Road, Beijing 100049, P.\ R.\ China}
\email{maixf@shao.ac.cn}

%% file: 1_intro.tex
High-mass stars ($M \geq 8~M_{\sun}$) typically form within dense regions of molecular clouds, known as hubs, where filaments converge \citep{2020A&A...642A.177D, 2020A&A...642A..87K, 2025A&A...699A.355Z}. 
These potential high-mass star-forming regions may exist as prestellar cores for $<10^4$~years \citep{2013A&A...558A.125D, 2017A&A...602A..77T, 2018ARA&A..56...41M}. 
During the early stages of high-mass star formation, stellar embryos with masses less than $8~M_{\sun}$ are IR-quiet protostars \citep{2007prpl.conf..165B, 2007A&A...476.1243M, 2018ARA&A..56...41M, 2010A&A...515A..55R}. 
Protostars become IR-bright when the stellar embryos reach $M \geq 8~M_{\sun}$, entering the high-mass protostellar object (HMPO) stage \citep{2007A&A...476.1243M, 2018ARA&A..56...41M, 2010A&A...515A..55R}. 
The HMPO stage can be further divided based on chemical characteristics into two categories: 
chemically poorer (earlier) HMPOs \citep{2014A&A...563A..97G} and later, hot molecular cores (HMCs) that exhibit molecular-rich chemistry \citep{2014A&A...563A..97G, 2021MNRAS.505.2801L, 2022MNRAS.511.3463Q}. 
HMCs are characterized by rich spectra of complex organic molecules (COMs), high temperatures ($\gtrsim100$~K), and compact source sizes ($\lesssim0.1$~pc) \citep{2005IAUS..227...59C}. 
Eventually, when the ultraviolet (UV) radiation from the embedded high-mass (proto)star ionizes the surrounding gas, a hyper-compact H\textsc{ii} (HCH\textsc{ii}) or ultra-compact H\textsc{ii} (UCH\textsc{ii}) region is formed. 
Due to the short evolutionary timescales, multiple stages can overlap; 
for instance, line-rich HMC chemistry and emission from ionized gas may coexist within the same core \citep{2021MNRAS.505.2801L}. 
Additionally, because high-mass stars typically form in clusters, individual cores at varying evolutionary stages are frequently observed in close proximity to one another \citep{2020MNRAS.496.2790L}. 

Around an accreting protostar, the centrifugal force of rotating material typically balances gravity, leading to the formation of a rotationally supported structure. 
While this often results in a Keplerian disk ($v \propto r^{-0.5}$, \citealp{2020SSRv..216...43Z}), actual disks frequently exhibit deviations from a strict Keplerian profile due to additional physical factors, such as self-gravity, internal gas pressure in hot disks, and magnetic braking effects. 
Some of the infalling material is ejected as a bipolar outflow, a process driven by complex magnetohydrodynamics (MHD) that helps carry away excess angular momentum \citep{2000ApJ...528L..41T}. 
This disk-outflow system is crucial for massive young stellar objects (MYSOs), as it protects the accretion process from the intense radiation ($L \geq 10^{3}~L_{\sun}$) pressure exerted by the central protostars \citep{2010ApJ...722.1556K}. 
Recently, the Atacama Large Millimeter/submillimeter Array (ALMA) has identified Keplerian disk candidates around several MYSOs \citep{2014A&A...571A..52B, 2015ApJ...813L..19J, 2016MNRAS.462.4386I, 2018ApJ...869L..24I, 2018ApJ...860..119G, 2019A&A...627L...6M, 2019A&A...622A.206M, 2022NatAs...6..837L, 2022MNRAS.509..748W}, some of which extend up to a few hundred au. 
These disks are often associated with COMs such as acetonitrile (\ce{CH3CN}, e.g., \citealp{2015ApJ...813L..19J, 2016MNRAS.462.4386I}) and methanol (\ce{CH3OH}, e.g., \citealp{2016MNRAS.462.4386I,2022MNRAS.509..748W}). 
In certain cases, refractory molecules like sodium chloride (NaCl, e.g., \citealp{2019ApJ...872...54G, 2023ApJ...942...66G, 2020ApJ...900L...2T}) or hydrogen recombination lines from ionized gas (e.g., \citealp{2023PASJ...75..225M}) can be used to trace the presence of these disks. 
 In contrast, bipolar molecular outflows are usually traced using carbon monoxide (CO, e.g., \citealp{2022A&A...658A.160Y}). 
 Silicon monoxide (SiO) line emissions frequently indicate central jet-like features along the axes of broader CO outflows (e.g., \citealp{2024ApJ...977..126L}). 

High-mass stars are often found in binary or multiple star systems. 
Multiple surveys, mainly in the optical and near-infrared regimes, have shown that the multiplicity fraction---defined as the proportion of primaries having at least one companion---among main-sequence high-mass stars exceeds 90\% \citep{2023ASPC..534..275O}. 
Despite this high percentage, the early formation phases of such massive binary systems remain largely unexplored. 
In low-mass binary systems, we can observe large rotating structures that surround both protostars and their accretion disks, commonly referred to as ``circumbinary disks", e.g., L1551 NE \citep{2012ApJ...754...52T, 2014ApJ...796....1T}, L1551 IRS 5 \citep{2014ApJ...796...70C}, [BHB 2007] 11 \citep{2019Sci...366...90A}. 
However, circumbinary disks have only been observed around a very limited number of massive proto-binary systems. 
One example is the massive protobinary system IRAS 16547--4247 in which 
a circumbinary disk was detected using the well-known disk tracer \ce{CH3CN} \citep{2020ApJ...900L...2T}, along with other COMs, such as methyl formate (\ce{CH3OCHO}, \citealp{2019ApJ...872..176Z}). 

MYSOs are rare and are predominantly found at distances exceeding 2~kpc. 
This distance poses significant challenges for observational studies due to limitations in spatial resolution. 
However, the deployment of the ALMA has made it possible to resolve the small structures surrounding these distant objects. 
One notable statistical survey focused on high-mass star-forming regions in the galactic plane is the ``ALMA Three-millimeter Observations of Massive Star-forming regions (ATOMS)'' project \citep{2020MNRAS.496.2790L}. 
The ATOMS project observed 146 massive protoclusters at a wavelength of 3~mm with resolutions ranging from 1\arcsec~to 2\arcsec. 
Additionally, 139 of these protoclusters were further observed at a wavelength of 1.3~mm by the ``Querying Underlying mechanisms of massive star formation with ALMA-Resolved gas Kinematics and Structures (QUARKS)'' project \citep{2024RAA....24b5009L}. 
The QUARKS project achieved high resolutions of $\sim$0\farcs3, while also covering large-scale flux up to 27\arcsec~through Atacama Compact 7-m Array (ACA) observations \citep{2024RAA....24f5011X}. 
The QUARKS frequency band captures numerous molecular tracers of disks and outflows (as detailed in Table 2 of \citealp{2024RAA....24b5009L}), making it particularly well-suited for investigating various aspects of disk-outflow systems surrounding MYSOs. 

IRAS 08303--4303 is one of the targets in the ATOMS and QUARKS projects, located at a distance of $2.35\pm0.55$~kpc \citep{2021MNRAS.505.2801L, 2024RAA....24b5009L}. 
\citet{2004A&A...426...97F} estimated the overall clump mass of this object to be $260~M_{\sun}$, based on 1.2~mm continuum flux measurements obtained with the Swedish-ESO submillimeter telescope.
They also derived the bolometric luminosity to be $\sim7000~L_{\sun}$ from the spectral energy distribution (SED) fitting, which incorporated both the 1.2~mm continuum flux and the fluxes from the four IRAS bands.  
\citet{2024RAA....24b5009L} determined the systemic LSR (Local Standard of Rest) velocity ($v_\mathrm{sys}$) of the overall clump to be approximately $14.5~\mathrm{km~s}^{-1}$, based on \ce{H^13CO^+} ATOMS data. 
The ATOMS project identified more than 20 emission lines towards the brightest core. 
Many of these emissions are from COMs \citep{2025A&A...697A.190L}. 
The calculated excitation temperature using \ce{CH3OH}, \ce{CH3OCHO}, and propionitrile (\ce{C2H5CN}) is $\sim$100~K, which supports the hot core chemistry \citep{2022MNRAS.511.3463Q}. 
It is worth noting that the \ce{H_{40\alpha}} hydrogen recombination line emission, which traces ionized H\textsc{ii} regions, was not detected. 
Consequently, \citet{2021MNRAS.505.2801L} classified it as a pure and strong hot molecular core candidate (pure s-cHMC). 
This classification is based on its high line richness ($N_\mathrm{line} \geq 20$) and the non-detection of the \ce{H_{40\alpha}} line. 
However, the ATOMS observations, with a resolution of $\sim$1\farcs5 (about 3500~au at a distance of 2.35~kpc), have not resolved the IRAS 08303--4303 region down to the scale of the disk. 

In this study, we investigate the physical and dynamical properties of dense condensations and disk-outflow systems in the IRAS 08303--4303 region by primarily analyzing the QUARKS data at a resolution of $\sim$0\farcs3 (about 700~au). 
This level of detail has enabled us to identify a candidate for a circumbinary disk within the aforementioned hot core. 
While recent higher-resolution ($\sim0\farcs05$) observations have further resolved the dense condensations down to the $\sim$100 au scale \citep{2026ApJ..1006...25Y}, our analysis provides a crucial link between the core-scale (ATOMS) and these small-scale structures. 
The observational details are summarized in Section~\ref{sec:obs}, while the analysis results are presented in Section~\ref{sec:result}. 
Section~\ref{sec:discuss} examines the kinematics of the circumbinary disk and its associated distinct features, and Section~\ref{sec:sum} summarizes the main findings of our research. 

%% file: 2_obs.tex
ALMA observations of the massive protostellar system IRAS 08303--4303 were conducted in Bands 3 and 6. 
The Band 3 observations took place on 2019 October 14 and 19, as part of the ATOMS program (Project ID: 2019.1.00685.S; PI: Tie Liu). 
The first observation was conducted with the ACA, and the second with the 12-m array in the C-4 configuration. 
Among the eight spectral windows (SPWs) used, the continuum image was generated from two wide SPWs, each with a bandwidth of 1875 MHz. 
The combined Band 3 continuum image from the ACA and 12-m array has a synthesized beam of 1\farcs55$\times$1\farcs16 with a position angle (PA) of $-$71\fdg3 and a noise level of $90~\mathrm{\mu Jy~beam}^{-1}$. 
Detailed information regarding the ATOMS observations can be found in \citet{2020MNRAS.496.2790L}. 

The Band 6 observations were carried out on 2021 November 6, 2022 December 29, and 2023 April 8, as part of the QUARKS program (Project IDs: 2021.1.00095.S, 2022.1.00298.S, and 2023.1.00425.S; PI: Lei Zhu). 
The observations utilized three configurations: the ACA and the 12-m array in C-3 and C-5 configurations. 
A comprehensive overview of each configuration's observation details is presented in Table~\ref{tab:obs}. 
The three configuration data sets were calibrated using the pipeline of the Common Astronomy Software Applications (CASA; \citealp{2022PASP..134k4501C}) and subsequently combined with the CASA task \texttt{concat} before imaging. 
For continuum imaging, we utilized all four 1875~MHz wideband spectral windows (SPWs) centered at 217.9, 220.3, 231.3, and 233.5~GHz, after manually masking the spectral lines. 
For both line and continuum imaging, we set the robust parameter to +0.5 in the CASA \texttt{tclean} task. 
Deconvolution was performed using the `multiscale' algorithm for line imaging and the `mtmfs' (Multi-Term Multi-Frequency Synthesis) algorithm for continuum imaging. 
Following this, we performed self-calibration in both phase and amplitude using the resulting 1.3~mm continuum image as a model, applying a solution interval equal to the length of a scan ($\sim60~\mathrm{sec}$). 
We then applied these self-calibration solutions to the spectral data. 
Finally, we generated line cubes for all four SPWs after subtracting the continuum using the CASA task \texttt{uvcontsub}. 
The combined Band 6 continuum image resulted in a synthesized beam of 0\farcs33$\times$0\farcs33 (PA = +66\fdg1) and a noise level of $0.15~\mathrm{mJy~beam}^{-1}$. 
All spectral line images have synthesized beams between 0\farcs3 and 0\farcs4, with a velocity resolution of $\sim1.3~\mathrm{km~s}^{-1}$ and a noise level of $3~\mathrm{mJy~beam}^{-1}$. 
These four SPWs were specifically designed to target various spectral lines, including \ce{CO}, \ce{SiO}, \ce{CH3OH}, and \ce{CH3CN}. 
Detailed properties and detections of these targeted lines are presented in Table~\ref{tab:line}. 
The maximum recoverable scale for this ALMA observation is 20\arcsec ($\sim$0.2 pc). 
More details about the QUARKS observations can be found in \citet{2024RAA....24b5009L}. 

\begin{deluxetable*}{cccccccc}[htbp]
\tablecaption{Observation details for each configuration, including observation date, baseline length (BL), angular resolution (AR), maximum recoverable scale (MRS), number of antennas (N Ant.), time on source (ToS), and employed calibrators. 
\label{tab:obs}}
\tablehead{
    \colhead {Configuration} & \colhead {Date} & \colhead {BL (m)} & 
    \colhead {AR (\arcsec)} & \colhead {MRS (\arcsec)} & 
    \colhead {N Ant.} & \colhead {ToS (min)} & \colhead {Calibrators}}
\startdata
ACA & 2021-11-06 & 8.9--45.0 & 4.9 & 20.0 & 9 & 5 & J1058+0133, J0922--3959\\
C-3 & 2022-12-29 & 15.1--500.6 & 0.7 & 7.3 & 46 & 1 & J1037--2934, J0922--3959\\
C-5 & 2023-04-08 & 15.3--1397.4 & 0.2 & 3.4 & 44 & 4 & J1107--4449, J0922--3959, J0849--3541\\
\enddata
\end{deluxetable*}

%% file: 3_1_cont.tex
\begin{figure*}[htbp]
    \centering
    \includegraphics[width=\textwidth]{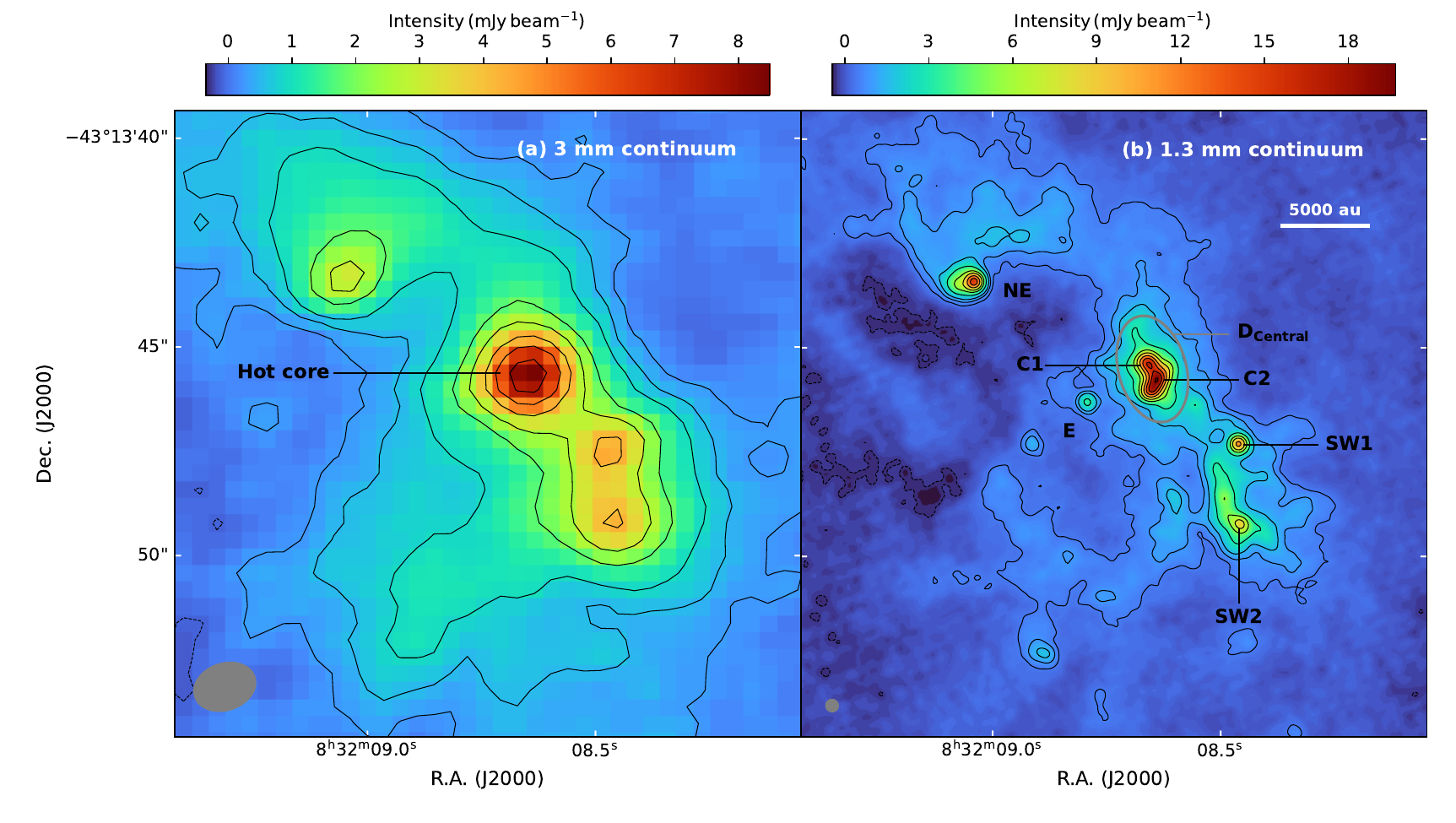}
    \caption{\textbf{(a)} ATOMS 3~mm dust continuum image without primary beam correction. 
    Contours are at [-1, 3, 6, 10, 20, 30, 45, 60, 80]$\sigma$, where $1\sigma = 90~\mathrm{\mu Jy~beam}^{-1}$. 
    \textbf{(b)} QUARKS 1.3~mm dust continuum image without primary beam correction. 
    Contours are at [-1.5, 3, 6, 10, 20, 35, 50, 70, 90, 110, 130]$\sigma$, where $1\sigma = 0.15~\mathrm{mJy~beam}^{-1}$. 
    In each panel, the gray ellipse in the lower-left corner represents the synthesized beam size. 
    \label{fig:cont}}
\end{figure*}

Figure~\ref{fig:cont} presents the continuum emission images of ATOMS and QUARKS. 
A central hot core in the lower-resolution 3~mm image \citep{2022MNRAS.511.3463Q} is resolved into two distinct condensations, labeled `C1' and `C2', when viewed in the higher-resolution 1.3~mm image. 
The projected distance between C1 and C2 is only 0\farcs38 ($\sim900~\mathrm{au}$), indicating that they likely form a binary system. 
Notably, C2 emits stronger continuum emission than C1 and is considered the primary component. 
Six condensations have been identified visually; 
five bright and dense condensations labeled NE, C1, C2, SW1, and SW2, and a faint and isolated condensation labeled E (Figure~\ref{fig:cont}(b)). 
All six condensations are aligned along a filamentary structure elongated in the northeast-southwest direction. 

\begin{figure*}[htbp]
\includegraphics[width=\textwidth]{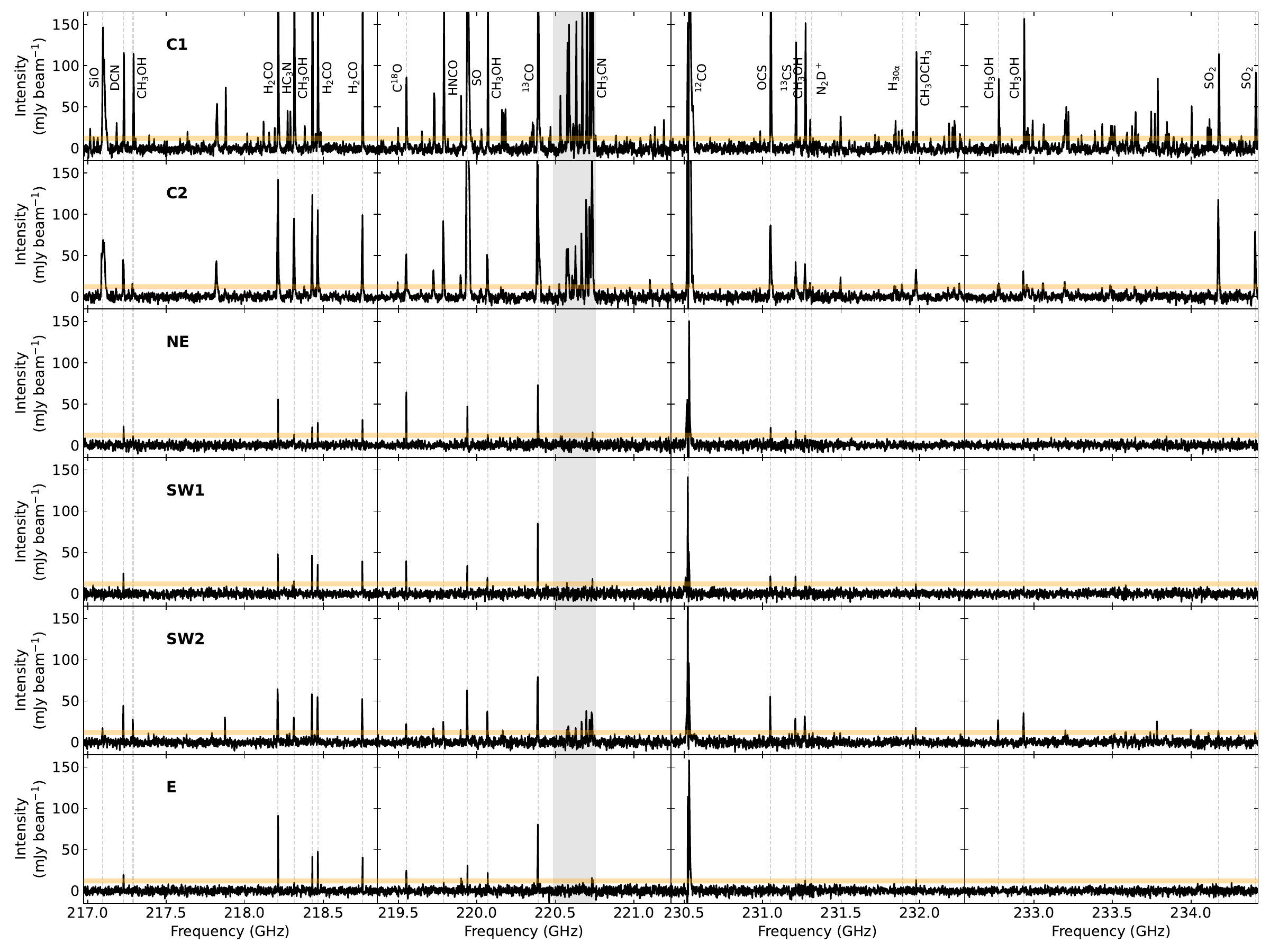}
\caption{
Spectra of the six dense condensations at the continuum peak. 
The yellow-filled area indicates the marginal detection regime ($3\sigma$--$5\sigma$) listed in Table~\ref{tab:line}. 
Vertical dotted lines mark line frequencies, with names labeled on them. 
For \ce{CH3CN} lines, multiple transitions occur within a narrow frequency range, so these are represented by filled areas instead. 
Note that \ce{H30_\alpha} is not detected in any condensation. }
\label{fig:spec}
\end{figure*}

Figure~\ref{fig:spec} presents the spectra extracted at the continuum peaks of the six dense condensations. Both C1 and C2 display line-rich HMC spectra, including multiple transitions of \ce{CH3CN} and \ce{CH3OH}. 
SW2 also shows these transitions, but their intensities are much weaker. 
In contrast, NE, SW1, and E show line-poor spectra, displaying only a few detectable lines, such as transitions of CO isotopologues and \ce{H2CO}, with only marginal detections otherwise. 
These transitions exhibit extended morphologies, which means that some may be affected by contamination from the nearby hot cores C1 and C2. 
This suggests that these condensations might be at a colder, earlier evolutionary stage than C1, C2, and SW2. 
Alternatively, they could be lower-mass objects. 
The NE condensation contains a stellar embryo at its center, as strong bipolar outflows have been detected (see Figure~\ref{fig:outflow}). 
Moreover, its outflow parameters are significantly larger than those of typical low-mass YSOs and comparable to those of MYSOs (see Section~\ref{subsec:result-outflow}). 
Thus, this line-poor signature implies that NE could be classified as a line-poor HMPO prior to the formation of a hot core, rather than as a low-mass YSO. 
However, there is no clear sign of star formation activity in the SW1 and E condensations. 

The \ce{H_{30\alpha}} line emission was not detected in any of the six condensations, indicating that none of them have reached the UCH\textsc{ii} phase. As mentioned above, the \ce{H_{40\alpha}} line emission was not detected, either \citep{2021MNRAS.505.2801L}. 
While an emission line was detected in C1 near the rest frequency of \ce{H_{30\alpha}}, its linewidth (full width at half maximum, FWHM), $3.5~ \mathrm{km}~\mathrm{s^{-1}}$, differs greatly from the broad linewidth of 25--30~$\mathrm{km}~\mathrm{s^{-1}}$ typically expected in UCH\textsc{ii} regions \citep{1989ApJS...69..831W, 2017A&A...602A..37K, 2018A&A...616A.107K}. 
Thus, it is likely one of the \ce{^{33}SO_2}, \ce{CH_3OCH_3}, or \ce{C2H5CN} lines, which have rest frequencies within 1~MHz of \ce{H_{30\alpha}}. 
However, since there is no cm-wavelength data available, the presence of an HC/UCH\textsc{ii} region cannot be completely ruled out. 
Table~\ref{tab:line} summarizes representative molecular lines and their detection toward each condensation. 
These lines were identified by comparing observed frequencies with those listed in the Cologne Database for Molecular Spectroscopy (CDMS, \citealp{2001A\string&A...370L..49M,2005JMoSt.742..215M, 2016JMoSp.327...95E}). 

\begin{deluxetable*}{llcllllll}[htbp]
\tablecaption{Representative molecular lines and their detection toward each condensation. 
\label{tab:line}}
\tablehead{
    \colhead {Line} & \colhead {Frequency\tablenotemark{a} (GHz)} & 
    \colhead {$E_{u}$\tablenotemark{a} (K)} & \colhead {NE} & 
    \colhead {C1} & \colhead {C2} & 
    \colhead {SW1\tablenotemark{e}} & \colhead {SW2} & \colhead {E\tablenotemark{e}}}
\startdata
\ce{CH3CN} $J_K=12_0\text{--}11_0$ & 220.7472617 & 68.9 & Y & Y & Y & Y & Y & Y\tablenotemark{c} \\
\ce{CH3CN} $J_K=12_1\text{--}11_1$ & 220.7430111 & 76.0 & n & Y & Y & Y\tablenotemark{c} & Y & Y \\
\ce{CH3CN} $J_K=12_2\text{--}11_2$ & 220.7302611 & 97.4 & n & Y & Y & n & Y & n \\
\ce{CH3CN} $J_K=12_3\text{--}11_3$ & 220.7090170 & 133.2 & Y\tablenotemark{c} & Y & Y & Y\tablenotemark{c} & Y & n \\
\ce{CH3CN} $J_K=12_4\text{--}11_4$ & 220.6792874 & 183.1 & n & Y & Y & n & Y & n \\
\ce{CH3CN} $J_K=12_52\text{--}11_5$ & 220.6410844 & 247.4 & n & Y & Y & n & Y & n \\
\ce{CH3CN} $J_K=12_6\text{--}11_6$ & 220.5944237 & 325.9 & n & Y & Y & n & Y & n \\
\ce{CH3CN} $J_K=12_7\text{--}11_7$ & 220.5393240 & 418.6 & n & Y & Y & n & n & n \\
\ce{CH3CN} $J_K=12_8\text{--}11_8$ & 220.4758078 & 525.6 & n & Y & Y\tablenotemark{c} & n & n & n \\
\ce{CO} $J=2\text{--}1$ & 230.5380000 & 16.6 & Y & Y & Y & Y & Y & Y \\
\ce{^13CO} $J=2\text{--}1$ & 220.3986842 & 15.9 & Y & Y & Y & Y & Y & Y \\
\ce{C^18O} $J=2\text{--}1$ & 219.5603577 & 15.8 & Y & Y & Y & Y & Y & Y \\
\ce{SiO} $J=5\text{--}4$ & 217.104919 & 31.3 & Y\tablenotemark{d} & Y & Y & n & Y & n \\
\ce{H_{30\alpha}} & 231.900928\tablenotemark{b} & -- & n & n & n & n & n & n \\
\ce{SO} $J_N=6_5\text{--}5_4$ & 219.94944 & 35.0 & Y & Y & Y & Y & Y & Y \\
\ce{SO2} $J_{K_a,K_c}=28_{3,25}\text{--}28_{2,26}$ & 234.187057 & 403.0 & n & Y & Y & n & Y\tablenotemark{c} & n \\
\ce{SO2} $J_{K_a,K_c}=16_{6,10}\text{--}17_{5,13}$ & 234.4215884 & 213.3 & n & Y & Y & n & Y\tablenotemark{c} & n \\
\ce{HC3N} $J=24\text{--}23$ & 218.32472 & 131.0 & Y\tablenotemark{c} & Y & Y & Y & Y & Y\tablenotemark{c} \\
\ce{H2CO} $J_{K_a,K_c}=3_{0,3}\text{--}2_{0,2}$ & 218.22219 & 21.0 & Y & Y & Y & Y & Y & Y \\
\ce{H2CO} $J_{K_a,K_c}=3_{2,2}\text{--}2_{2,1}$ & 218.47563 & 68.1 & Y & Y & Y & Y & Y & Y \\
\ce{H2CO} $J_{K_a,K_c}=3_{2,1}\text{--}2_{2,0}$ & 218.76007 & 68.1 & Y & Y & Y & Y & Y & Y \\
\ce{N2D+} $J=3\text{--}2$ & 231.321828 & 22.2 & n & n & n & n & n & n \\
\ce{CH3OH} $J_{K_a,K_c}=4_{2,3}\text{--}3_{1,2}~E$ & 218.44006 & 45.5 & Y & Y & Y & Y & Y & Y \\
\ce{CH3OH} $J_{K_a,K_c}=8_{0,8}\text{--}7_{1,6}~E$ & 220.078561 & 96.6 & Y\tablenotemark{c} & Y & Y & Y & Y & Y \\
\ce{CH3OH} $J_{K_a,K_c}=10_{2,9}\text{--}9_{3,6}~A^-$ & 231.28111 & 165.3 & Y\tablenotemark{c} & Y & Y & Y\tablenotemark{c} & Y & Y\tablenotemark{c} \\
\ce{CH3OH} $J_{K_a,K_c}=10_{-3,7}\text{--}11_{-2,9}~E$ & 232.94580 & 190.4 & n & Y & Y & n & Y & n \\
\ce{CH3OH} $J_{K_a,K_c}=6_{1,5}\text{--}7_{2,5}~A^-,~v_t=1$ & 217.29921 & 373.9 & Y\tablenotemark{c} & Y & Y & n & Y & n \\
\ce{CH3OH} $J_{K_a,K_c}=18_{3,16}\text{--}17_{4,13}~A^+$ & 232.78345 & 446.5 & n & Y & Y & n & Y & n \\
\ce{DCN} $J=3\text{--}2$ & 217.2385378 & 20.9 & Y & Y & Y & Y & Y & Y \\
\ce{H_2^13CO} $J_{K_a,K_c}=3_{1,2}\text{--}2_{1,1}$ & 219.90852 & 32.9 & n & Y & Y & n & Y & Y \\
\ce{HNCO} $J_{K_a,K_c}=10_{0,10}\text{--}9_{0,9}$ & 219.798274 & 58.0 & n & Y & Y & n & Y & Y\tablenotemark{c} \\
\ce{CH3OCH3} $J_{K_a,K_c}=13_{0,13}\text{--}12_{1,12}$ & 231.987858\tablenotemark{f} & 80.9 & n & Y & Y & Y\tablenotemark{c} & Y & Y\tablenotemark{c} \\
\ce{^13CS} $J=5\text{--}4$ & 231.220685 & 33.3 & Y & Y & Y & Y & Y & n \\
\ce{OCS} $J=19\text{--}18$ & 231.060993 & 110.9 & Y & Y & Y & Y & Y & n \\
\enddata
\tablenotetext{a}{From CDMS Spectral Line Catalog \citep{2001A\string&A...370L..49M,2005JMoSt.742..215M, 2016JMoSp.327...95E}}
\tablenotetext{b}{Data from the Splatalogue database of recombination lines}
\tablenotetext{c}{Marginally detected. Peak intensity: 9--15~$\mathrm{mJy~beam}^{-1}$ ($3\sigma$--$5\sigma$)}
\tablenotetext{d}{Not detected at the continuum peak but detected in the redshifted outflow region of this YSO}
\tablenotetext{e}{The spectrum may be contaminated by the outflow lobe from the central protobinary system. }
\tablenotetext{f}{Four transitions corresponding to the AA, EE, AE, and EA substates are blended; 
the reference frequency of the EE substate is given as representative. }
\end{deluxetable*}

\begin{deluxetable*}{ccccccccc}[htbp]
\tabletypesize{\footnotesize}
\tablewidth{0pt}
\tablecaption{
Physical properties of the six continuum condensations in 1.3~mm. 
\label{tab:cont_prop}}
\tablehead{
    \colhead {Name} & \multicolumn{2}{c}{Center (J2000)} & 
    \colhead {$\mathrm{FWHM}$\tablenotemark{a}} & {PA\tablenotemark{a}} & 
    \colhead {$I_{\mathrm{peak}}$} & \colhead {$F_{\mathrm{1.3~mm}}$} & 
    \colhead {Temperature} & \colhead {Mass}\\
    \colhead {} & \colhead {R.A.} & \colhead {Dec.} & 
    \colhead {($\arcsec\times\arcsec$)} & \colhead {(\arcdeg)} & 
    \colhead {($\mathrm{mJy~beam}^{-1}$)} & \colhead {(mJy)} & 
    \colhead {(K)} & \colhead {($M_{\sun}$)}}
\startdata
NE & 08\rah32\ram09\fs05 & -43\arcdeg13\arcmin43\farcs4 & $0.50\pm0.02\times0.32\pm0.02$ & 
$95\pm4$ & $12.6\pm0.3$ & $31.8\pm1.0$ & $75 \pm 14$ & $0.9\pm0.5$\\
C1 & 08\rah32\ram08\fs66 & -43\arcdeg13\arcmin45\farcs4 & $0.30\pm0.01\times0.22\pm0.01$ & 
$60\pm7$ & $14.9\pm0.3$ & $24.4\pm0.7$ & $235\pm3$\tablenotemark{b} & $0.2\pm0.1$\\
C2 & 08\rah32\ram08\fs64 & -43\arcdeg13\arcmin45\farcs9 & $0.55\pm0.01\times0.22\pm0.01$ & 
$147\pm1$ & $19.6\pm0.3$ & $46.0\pm0.9$ & $171\pm6$\tablenotemark{b} & $0.5\pm0.3$\\
SW1 & 08\rah32\ram08\fs46 & -43\arcdeg13\arcmin47\farcs3 & $0.11\pm0.04\times0.08\pm0.04$ & 
$166\pm115$ & $13.3\pm0.3$ & $14.4\pm0.5$ & $38 \pm 7$ & $0.8\pm0.5$\\
SW2 & 08\rah32\ram08\fs45 & -43\arcdeg13\arcmin49\farcs3 & $0.69\pm0.05\times0.21\pm0.03$ & 
$67\pm2$ & $6.4\pm0.3$ & $17.2\pm1.0$ & $262\pm9$\tablenotemark{b} & $0.13\pm0.06$\\
E & 08\rah32\ram08\fs79 & -43\arcdeg13\arcmin46\farcs3 & $0.28\pm0.07\times0.22\pm0.08$ & 
$57\pm40$ & $3.2\pm0.3$ & $5.1\pm0.7$ & $38 \pm 7$ & $0.3\pm0.2$\\
\hline
\multicolumn{9}{c}{Diffuse Components}\\
\hline
 & 08\rah32\ram08\fs97 & -43\arcdeg13\arcmin41\farcs9 & $6.2\pm0.3\times2.3\pm0.1$ & 
$80\pm2$ & $1.29\pm0.07$ & $170\pm9$ & \\
Central & 08\rah32\ram08\fs65 & -43\arcdeg13\arcmin45\farcs5 & $2.59\pm0.09\times1.59\pm0.06$ & 
$16\pm3$ & $3.5\pm0.1$ & $136\pm5$ & $127\pm2$\tablenotemark{b} & $2.1\pm1.0$\\
 & 08\rah32\ram08\fs42 & -43\arcdeg13\arcmin48\farcs5 & $4.9\pm0.3\times1.5\pm0.1$ & 
$31\pm2$ & $1.5\pm0.1$ & $106\pm6$ & \\
 & 08\rah32\ram08\fs49 & -43\arcdeg13\arcmin48\farcs8 & $1.80\pm0.10\times0.36\pm0.03$ & 
$14\pm1$ & $4.5\pm0.2$ & $36\pm2$ & \\
 & 08\rah32\ram08\fs65 & -43\arcdeg13\arcmin49\farcs5 & $6.9\pm0.4\times3.3\pm0.2$ & 
$111\pm3$ & $0.95\pm0.05$ & $196\pm11$ & \\
\enddata
\tablenotetext{a}{Deconvolved quantities}
\tablenotetext{b}{Temperature was calculated using \ce{CH3OH} (see Appendix~\ref{sec:appendixA})}
\end{deluxetable*}

The source position, deconvolved size (FWHM), position angle (PA), peak intensity, and integrated flux density of six dense condensations were derived from two-dimensional Gaussian fits using the CASA task \texttt{imfit}. 
Five additional diffuse, extended components were also fitted along with the dense condensations to perform proper fittings around bright peaks after subtracting the surrounding diffuse emission. 
The temperatures of the C1, C2, SW2 condensations and the central diffuse component were determined from the rotation diagram analyses of multiple \ce{CH3OH} transitions, assuming the case of optically thin line emission in the local thermal equilibrium (LTE) condition (see Appendix~\ref{sec:appendixA} for details). 
The dust temperature ($T_\mathrm{d}$) was assumed to be equal to the derived rotational temperature. 
For the other condensations,  the \ce{CH3OH} transitions with $E_u > 100~\mathrm{K}$ were not significantly detected, so the temperatures cannot be calculated from rotational diagram analysis. 
Three formaldehyde (\ce{H2CO}) transitions were detected in all dense condensations, but the \ce{H2CO} $3_{0,3}\text{--}2_{0,2}$ line is too optically thick to apply the method of \citet{2017A&A...598A..30T, 2018A&A...609A..16T}. 

For the NE condensation, the line-poor spectrum and strong bipolar outflow suggest that it is a line-poor HMPO. 
The typical temperature for such an object is 50~K \citep{2014A&A...563A..97G}, but the continuum condensation observed in this study has a much smaller scale than the ones detected in previous single-dish observations, so the temperature could be higher. 
However, it is expected to be lower than the typical hot core temperature of $\gtrsim 100$~K, giving a range of 50--100~K. 
Assuming a uniform distribution of values between 50 and 100~K, this can be represented as $75 \pm 14$~K (the standard deviation of a uniform distribution on $[a,b]$ is $(b-a)/\sqrt{12}$). 
For the SW1 and E condensations, star formation activity is unclear, and the temperature is likely lower than the 50~K typical of HMPOs. 
However, it might be higher than about 25~K because \ce{N2D+} was not detected toward either condensation. 
\ce{N2D+} is known to be depleted above the CO sublimation temperature ($\sim$25~K) due to reactions with CO and electrons \citep{2022ApJ...939..102L, 2022ApJ...925..144S}. 
Similarly to NE, assuming a uniform distribution of temperatures between 25 and 50~K, this can be represented as $38 \pm 7$ K. 

The masses of the condensations were calculated using the following equation \citep{2021MNRAS.505.2801L} assuming that the 1.3~mm continuum is optically thin dust emission, 
\begin{equation}
M = \frac{d^2 F_{\nu} R_\mathrm{gd}}{B_{\nu}(T_\mathrm{d}) \kappa_{\nu}}.
\end{equation} 
Here, $d$ is the distance to the source, $F_{\nu}$ is the integrated flux density (at $\nu=225.7~\mathrm{GHz}$) over the source size,
$R_\mathrm{gd}$ is the gas-to-dust mass ratio (assumed to be 100), 
$B_{\nu}(T_\mathrm{d})$ is the Planck function at temperature $T_\mathrm{d}$,
and $\kappa_{\nu}$ is the dust absorption coefficient, assumed to be $0.9~\mathrm{cm}^{2}~\mathrm{g}^{-1}$ at 1.3~mm. 
This value corresponds to grains covered by a thin ice mantle and coagulated at gas densities of $10^6~\mathrm{cm}^{-3}$ over a coagulation timescale of $10^5$ years \citep{1994A&A...291..943O}. 
We adopt uncertainties of 23\% for $R_\mathrm{gd}$ and 28\% for $\kappa_{\nu}$ in the calculation \citep{2017ApJ...841...97S, 2019ApJ...886..102S}. 

All the dense condensations have masses of $M \lesssim 1~M_{\sun}$ (Table~\ref{tab:cont_prop}). 
It should be noted that these are not the protostellar masses but the masses of the surrounding dense gases; 
much of the mass may have already been accreted onto the central objects. 
Additionally, a significant fraction of the cold, lower-density gas may not be observed owing to sensitivity limits or being resolved out, and therefore not included in the mass calculation. 
There is also a diffuse continuum component surrounding the central binary system. 
Using the same assumptions as before, the temperature of this component is estimated to be $127\pm2~\mathrm{K}$ based on \ce{CH3OH}, which corresponds to a component mass of $\sim2~M_{\sun}$. 
This component may represent the circumbinary disk around the protobinary system (see Sections~\ref{subsec:result-disk} and~\ref{subsec:discuss-disk} for further details). 
The observed properties of the six dense condensations and the surrounding diffuse components are summarized in Table~\ref{tab:cont_prop}. 

%% file: 3_2_outflow.tex
\begin{figure*}[htbp]
    \centering
    \includegraphics[width=\textwidth]{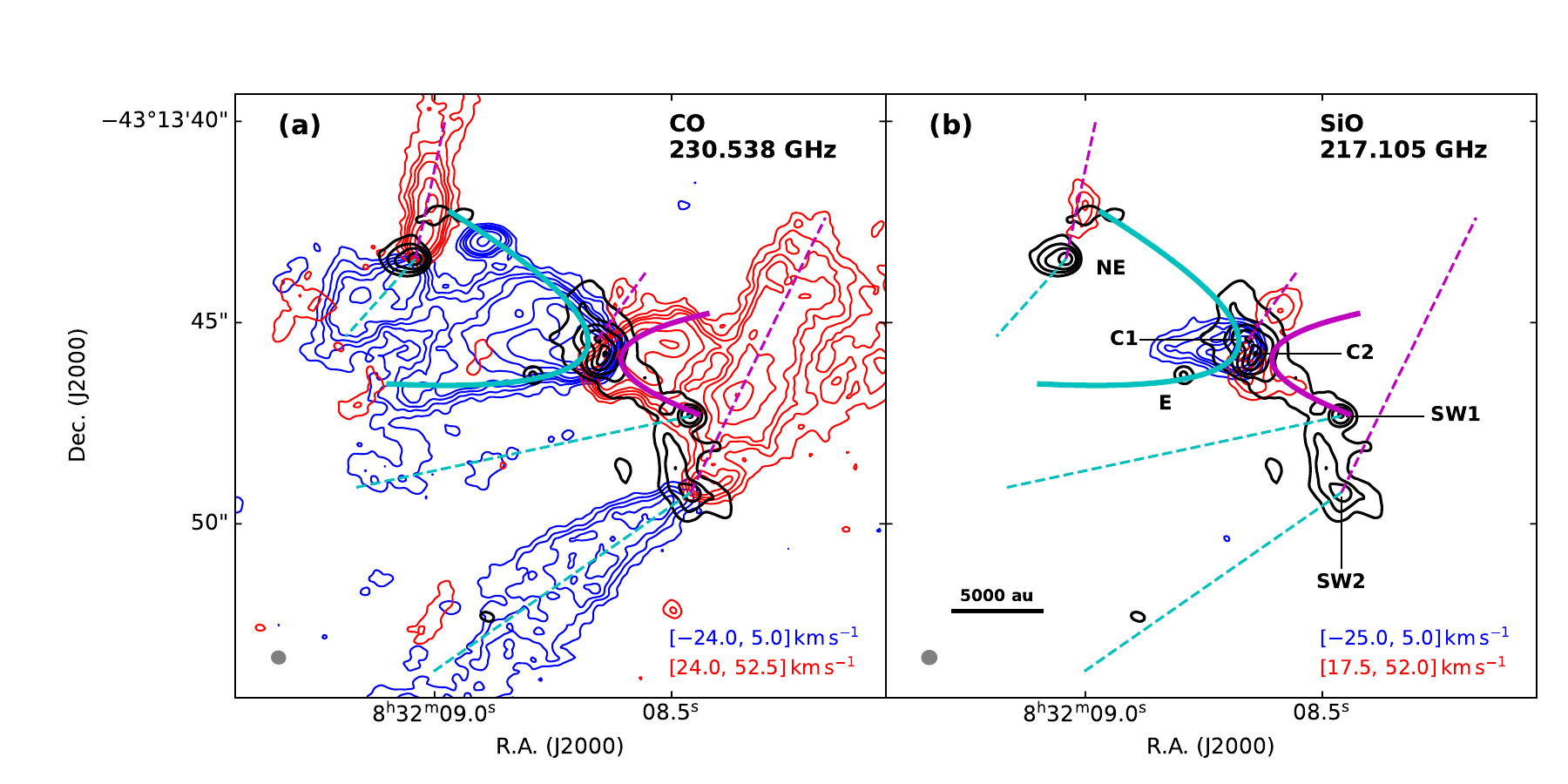}
    \caption{
    \textbf{(a)} The CO $J=2\text{--}1$ line integrated intensity map. 
    \textbf{(b)} The SiO $J=5\text{--}4$ line integrated intensity map. 
    In both panels, blue and red contours depict the distributions of blue- and redshifted gas, with contour levels at [5, 10, 15, 20, 30, 50, 70, 100]$\sigma$. 
    Black contours represent the 1.3~mm continuum emission distribution, with contour levels at [10, 20, 35, 50, 70, 90, 110, 130]$\sigma$. 
    All continuum contours in this figure use the same levels as in subsequent figures. 
    The integrated velocity ranges are presented in the bottom-right corner. 
    The gray ellipse in the bottom-left corner indicates the synthesized beam size. 
    The cyan and magenta lines/parabolas show the directions of the candidate outflow lobes. 
    \label{fig:outflow}}
\end{figure*}

Figure~\ref{fig:outflow}(a) displays the outflow image of CO ($J=2\text{--}1$) line emission. 
At least three distinct bipolar outflows are observed, originating from the NE and SW2 condensations, as well as the central protobinary system (C1 and C2). 
The SW1 and E condensations do not exhibit clear outflow features. 
However, weak blueshifted CO emission near SW1 may indicate a tentative outflow. 
The redshifted counterpart of the tentative SW1 outflow is difficult to identify due to blending with nearby redshifted outflows, and any detection remains highly uncertain. 
For the central bipolar outflow, it is challenging to identify the driving source owing to the close separation between C1 and C2. 
However, it is more likely to originate from C2, the primary component, as the continuum position angle of C2, $147\pm1\arcdeg$, is nearly perpendicular to the outflow direction. 
In this case, the weaker redshifted outflow lobe to the northwest of C1 may represent the outflow from the C1 condensation; 
its blueshifted counterpart would then overlap with the blueshifted lobe of C2’s outflow. 
If so, this NW--SE outflow is nearly perpendicular to the continuum PA of C1, $60\pm7\arcdeg$. 

Figure~\ref{fig:outflow}(b) shows the outflow image of SiO ($J=5\text{--}4$) line emission. 
Similar to CO, SiO is also a tracer of molecular outflows from MYSOs \citep{2021ApJ...921...96L, 2023A&A...677A.148G}. 
SiO often traces central jet-like features along the axes of broader CO outflows (e.g., \citealp{2024ApJ...977..126L}). 
Consistent with this expectation, the detected blueshifted SiO emission ridge is located near the center of the CO lobe, and its extension intersects with condensation C2. 
This supports the earlier interpretation that the central outflow originates from the primary component, C2. 
There is also redshifted SiO emission corresponding to the redshifted CO lobe northwest of C1. 
However, its blueshifted counterpart is blended with the outflow emission from C2; 
therefore, higher angular resolution observations are required to disentangle these components and confirm the presence of an outflow originating from C1. 

We derived the physical and dynamical parameters of the three outflows using a similar approach to that of \citet{2002A&A...383..892B} 
(see also \citealt{1990ApJ...348..530C, 2006ApJ...643..978K}). 
We assumed that the excitation temperature of CO gas is 30~K \citep{2002A&A...383..892B}, and the line intensity ratio of \ce{^12CO} to \ce{^13CO} in the line wing velocity range is 10 \citep{1993ApJ...417..624C}. 
The \ce{^13CO} ($J=2\text{--}1$) line was also covered in the QUARKS frequency band, but for this target, the line emission is weak in the high-velocity wings, making it difficult to calculate a reliable line intensity ratio. 
Therefore, we adopted a typical literature value instead. 
The assumed fractional abundance of \ce{^13CO} is $[\mathrm{H_2}]/[\mathrm{^{13}CO}] = 8.9 \times 10^5$ \citep{1992A&A...261..274C}, and the mean molecular weight is 2.8. 
For simplicity, we took the statistical mean value of the inclination angle ($i$), assuming that the outflows are randomly oriented. 
For a random distribution of orientations, the statistical mean inclination angle is not 45\arcdeg~but $\sim$57\fdg3 \citep{1996A&A...311..858B}, because the probability distribution is weighted by $\sin i$, giving greater weight to nearly edge-on orientations. 
We first derived the maximum three-dimensional velocity of each outflow lobe, $v_\mathrm{max}$, defined as
$(|v_\mathrm{LSR}-v_\mathrm{sys}|/\cos i)_\mathrm{max}$, where $v_\mathrm{LSR}$ is the observed LSR velocity (see Table~\ref{tab:outflow}). 
Dynamical parameters, including momentum, energy, and mechanical force, were calculated following the $v_{\mathrm{max}}$ method described in Section 4.2.1 of \citet{2013A&A...556A..76V}. 
This approach assumes that each outflow lobe moves at a single characteristic bulk velocity $v_\mathrm{max}$. 
Although this method may underestimate the outflow force by a factor of 1.5--3 compared to alternative techniques discussed in \citet{2013A&A...556A..76V}, it remains one of the most widely adopted approximations in the literature \citep{1992A&A...261..274C, 1998ApJ...502..315H, 2007A&A...471..873D}. 
\citet{2014MNRAS.444..566D} summarize in their Table~4 how these dynamical parameters are affected by the inclination angle. 
The accretion rate was estimated based on the conservation of momentum between the jet and the outflow. 
Due to the non-detection of a high-velocity SiO component (Figure~\ref{fig:outflow}), which precluded a direct measurement of the jet velocity, we adopted a representative value of $v_{\mathrm{jet}} = 500~\mathrm{km}~\mathrm{s^{-1}}$ \citep{1998ApJ...502..337M}. 
This approach differs from that of \citet{2002A&A...383..892B}, which assumed a fixed velocity ratio of $v_{\mathrm{jet}}/v_{\mathrm{out}} = 20$ based on the averaged outflow velocity ($v_{\mathrm{out}} \sim 28~\mathrm{km}~\mathrm{s^{-1}}$) derived from their single-dish data. 
Since our ALMA observations resolve much higher outflow velocities ($v_{\mathrm{out}} > 100~\mathrm{km}~\mathrm{s^{-1}}$), adopting this constant ratio would result in an unphysically large $v_{\mathrm{jet}}$, leading to an underestimation of ${\dot{M}}_\mathrm{jet}$ and ${\dot{M}}_\mathrm{acc}$.
We therefore employed a fixed $v_{\mathrm{jet}}$ consistent with typical high-mass protostellar jets \citep{1998ApJ...502..337M}. 
Additionally, we assumed a ratio of jet mass-loss rate to accretion rate (${\dot{M}}_\mathrm{jet}/{\dot{M}}_\mathrm{acc}$) of approximately 0.3 \citep{1998ApJ...502L.163T, 1999ASIC..540..193S}. 
Due to the various assumptions, the resulting outflow parameters have large uncertainties: a factor of $\sim3$ for the outflow mass, a factor of $\sim10$ for the mechanical force, and a factor of $\sim30$ for the mechanical luminosity \citep{2014MNRAS.444..566D}. 
Table~\ref{tab:outflow} lists the derived parameters. 
Our estimates are consistent with those of MYSOs with $\dot{M}_{\mathrm{out}} \geq 10^{-4}~M_{\sun}~\mathrm{yr}^{-1}$ \citep{2018ApJS..235....3Y, 2021MNRAS.507.4316B}, but they are significantly larger than those of low-mass YSOs with $\dot{M}_{\mathrm{out}} \leq 10^{-6}~M_{\sun}~\mathrm{yr}^{-1}$ \citep{2021A&A...648A..41V, 2024AJ....167...72D}, indicating that the driving sources are most likely MYSOs. 

\begin{deluxetable*}{cccccccccccccccc}[htbp]
\tablecaption{
Outflow Parameters: 
integrated velocity ranges of blue- and redshifted lobes $v_{\mathrm{LSR},b}$, $v_{\mathrm{LSR},r}$ 
[$\mathrm{km~s}^{-1}$]; 
average \ce{H2} column density at the outflow lobe $N_b$, $N_r$ [$10^{21}~\mathrm{cm}^{-2}$]; 
masses $M_b$, $M_r$ and $M_\mathrm{out}$ ($M_b + M_r$) [$10^{-2}~M_{\sun}$]; 
momentum $p$ [$M_{\sun}~\mathrm{km~s}^{-1}$]; 
energy $E$ [$10^{45}~\mathrm{erg}$]; 
average length of the outflow lobes $r$ [$10^3~\mathrm{au}$]; 
dynamical time $t_\mathrm{dyn}$ [yr]; 
outflow rate ${\dot{M}}_\mathrm{out}$ [$10^{-4}~M_{\sun}~\mathrm{yr}^{-1}$]; 
accretion rate ${\dot{M}}_\mathrm{acc}$ [$10^{-4}~M_{\sun}~\mathrm{yr}^{-1}$]; 
mechanical force $F_m$ [$10^{-2}~M_{\sun}~\mathrm{km~s}^{-1}~\mathrm{yr}^{-1}$]; 
and mechanical luminosity $L_m$ [$10^2~L_{\sun}$]. 
All parameters but the velocities were corrected for the inclination $i~\left(57\fdg3\right)$. 
The uncertainties of the resulting outflow parameters are a factor of $\sim3$ for the outflow mass $M_\mathrm{out}$, $\sim10$ for the mechanical force $F_m$, and $\sim30$ for the mechanical luminosity $L_m$. 
\label{tab:outflow}}
\tablehead{
    \colhead {} & $v_{\mathrm{LSR},b}$\tablenotemark{a} & $v_{\mathrm{LSR},r}$\tablenotemark{a} & 
    \colhead {$N_b$} & \colhead {$N_r$} & 
    \colhead {$M_b$} & \colhead {$M_r$} & \colhead {$M_\mathrm{out}$} & 
    \colhead {$p$} & \colhead {$E$} & \colhead {r} & 
    \colhead {$t$} & \colhead {$\dot{M}_\mathrm{out}$} & \colhead {$\dot{M}_\mathrm{acc}$} & 
    \colhead {$F_m$} & \colhead {$L_m$}}
\startdata
NE & [-56.2, 9.2] & [18.7, 62.5] & 3.02 & 3.23 & 4.88 & 2.03 & 6.91 & 8.15 & 9.77 & 8.50 & 367 & 1.88 & 1.38 & 2.22 & 2.21\\
C & [-68.3, 9.2] & [18.7, 55.6] & 5.25 & 6.16 & 19.20 & 10.63 & 29.83 & 37.37 & 50.31 & 9.39 & 388 & 7.68 & 5.87 & 9.62 & 10.7\\
SW2 & [-59.4, 9.2] & [18.7, 49.2] & 3.34 & 4.67 & 7.42 & 31.46 & 38.87 & 29.68 & 26.17 & 19.90 & 939 & 4.14 & 2.77 & 3.16 & 2.31\\
\enddata
\tablenotetext{a}{The velocity range in Figure~\ref{fig:outflow} is set differently for better visualization. }
\end{deluxetable*}

%% file: 3_3_diskC1C2.tex
\begin{figure*}[htbp]
\centering
\includegraphics[width=\textwidth]{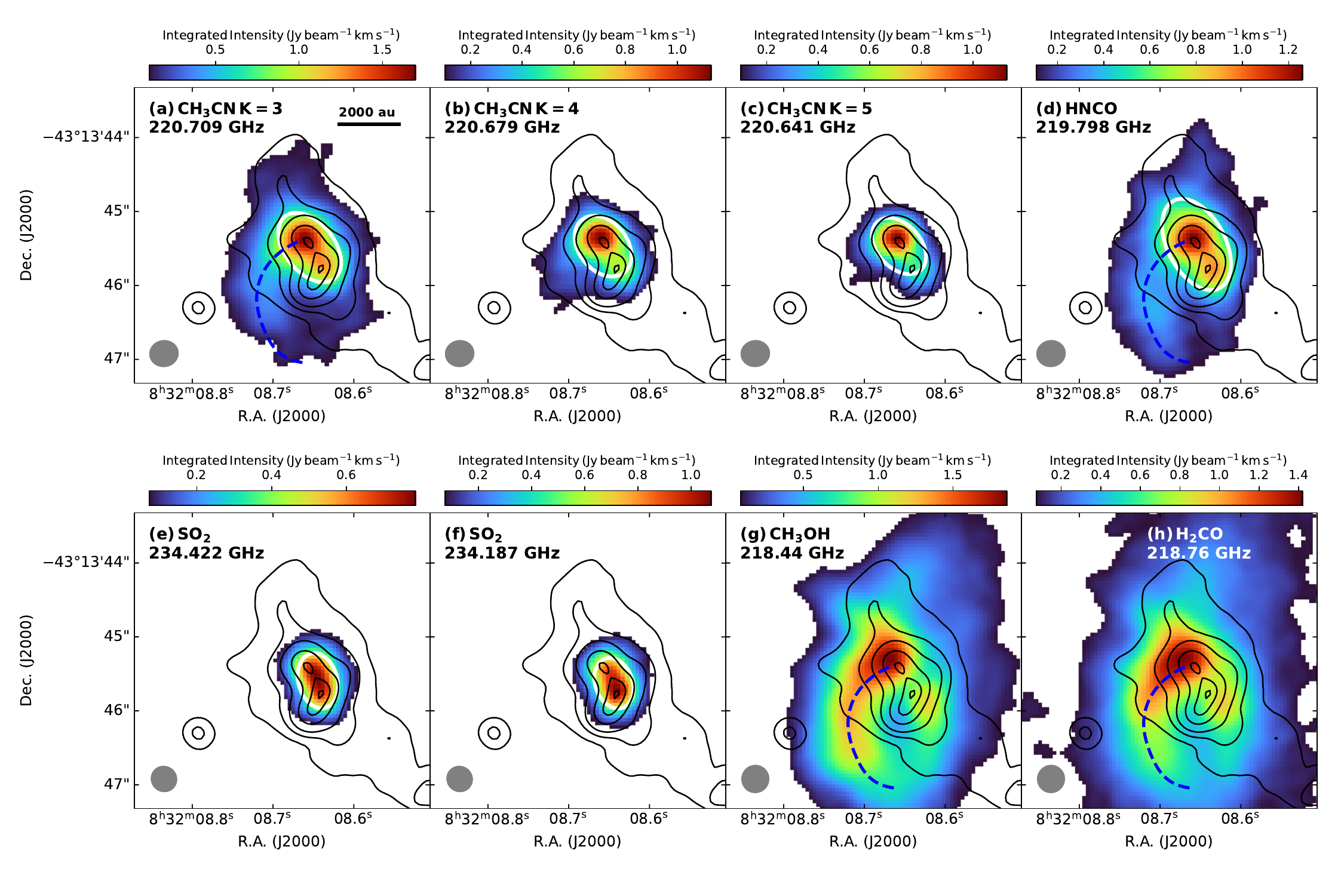}
\caption{
Integrated intensity maps (Moment 0) of 8 molecular transitions around C1 and C2 (color). 
All images are masked below $5\sigma$. 
Black contours represent the distribution of 1.3~mm continuum emission. 
In panels (a)--(f), the open white ellipse near the peak of the emission shows the 2D Gaussian fitting result of the integrated intensity map, which corresponds to the FWHM of the circumbinary disk. 
In panels (a), (d), (g), and (h), the blue dashed line marks the candidate streamer, tracing the emission ridge in the \ce{CH3OH} integrated intensity map. 
The filled gray ellipse in the lower-left corner indicates the size of the synthesized beam. 
\label{fig:mom0}}
\end{figure*}

\begin{figure*}[htbp]
\centering
\includegraphics[width=\textwidth]{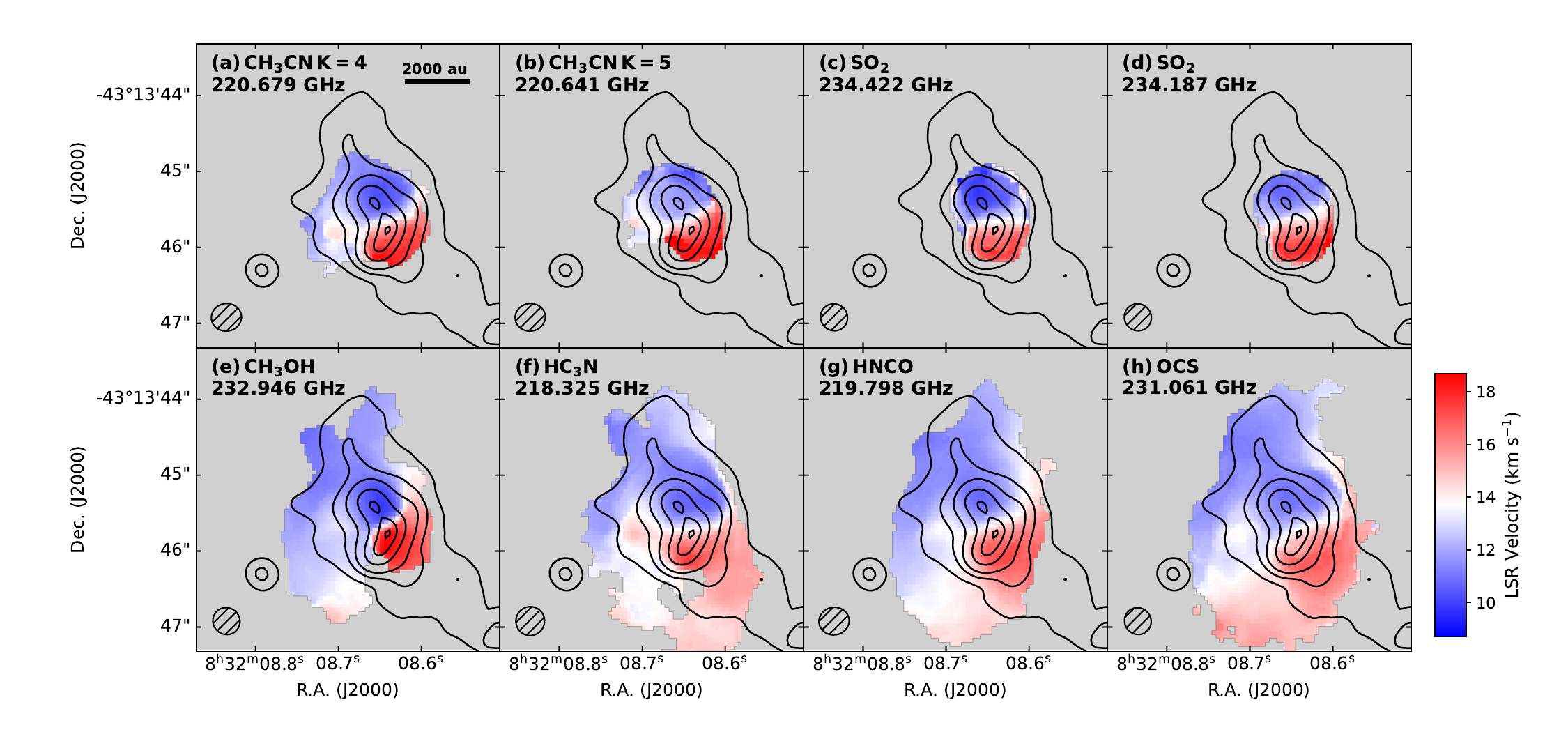}
\caption{
Intensity-weighted velocity maps (Moment 1) around C1 and C2 for 8 molecular transitions (color). 
All images are masked below $5\sigma$. 
Black contours represent the distribution of 1.3~mm continuum emission. 
The systemic velocity of 13.7~$\mathrm{km}~\mathrm{s^{-1}}$ was determined by fitting the 234.187~GHz \ce{SO2} transition, which shows similar intensities between the two continuum peaks. 
The black hatched ellipse in the lower-left corner indicates the size of the synthesized beam. 
\label{fig:mom1}}
\end{figure*}

The \ce{CH3CN} molecular lines have frequently been used to observe accretion disks around MYSOs (e.g., \citealt{2015ApJ...813L..19J, 2018ApJ...869L..24I}). 
The line emission was detected toward all the six continuum condensations, 
although it was only marginally detected in the E condensation (Table~\ref{tab:line}). 
Figure~\ref{fig:mom0} shows the integrated line intensity maps of several molecular lines, including \ce{CH3CN}, HNCO, and \ce{SO2}, which trace an elongated structure surrounding the C1 and C2 condensations. 
\citet{2016MNRAS.462.4386I} found that molecules such as \ce{CH3OH} and \ce{H2CO} also trace the circumstellar disk in the case of the MYSO G11.92--0.61 MM1, similar to \ce{CH3CN}. 
Figure~\ref{fig:mom1} suggests that this elongated structure may represent a circumbinary disk, as a consistent velocity gradient is observed along its  major axis in multiple molecular lines. 
This finding is similar to the ALMA observational results of IRAS 16547--4247 in which \ce{CH3CN} traces the circumbinary disk of a high-mass protobinary system \citep{2020ApJ...900L...2T}. 
Also, the higher-velocity component relative to the systemic velocity is located near the center, consistent with Keplerian rotation. 
Additionally, enhanced \ce{CH3CN} velocity dispersion is largely confined to the extent of the proposed circumbinary disk, apart from a localized enhancement immediately west of the C1/C2 center that may indicate a minor outflow contribution (see Appendix~\ref{sec:appendixB}). 
Assuming that the circumbinary disk is circular, we derived its inclination angle from the six transitions shown in Figure~\ref{fig:mom0}~(a)--(f) as $\cos^{-1}(D_{\mathrm{min}}/D_{\mathrm{maj}}) = 55 \pm 3\arcdeg$, where $D_{\mathrm{min}}$ and $D_{\mathrm{maj}}$ are the minor and major diameters, respectively. 
The small uncertainty displays that the six measurements agree well. 
Interestingly, the majority of the detected molecular lines, with the exception of the \ce{SO2} lines, are significantly brighter in C1 than in C2, which is opposite to what is observed in the continuum emission. 
Another interesting aspect of our findings is that \ce{CH3OH}, \ce{H2CO}, \ce{CH3CN}, and HNCO exhibit an arc-like structure extending southeast from C1. 
This feature may represent a streamer, similar to those observed in the high-mass protostellar core G336.01--0.82 \citep{2023ApJ...959L..31O, 2025SciA...11.4512O}. 
We discuss this feature further in Section~\ref{subsec:discuss-streamer}. 

%% file: 3_4_diskSW2.tex
\begin{figure}[!ht]
\centering
\includegraphics[width=0.45\textwidth]{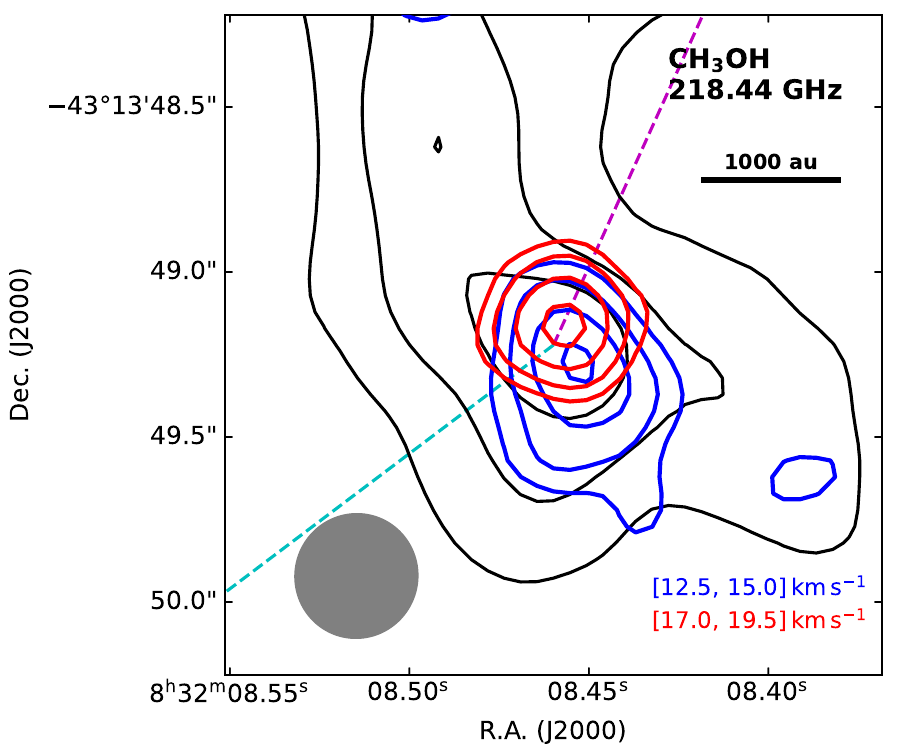}
\caption{\ce{CH3OH} $4_{2,3}\text{--}3_{1,2}~E$ line integrated intensity map around the SW2 condensation, with contour levels at [7, 9, 12, 15]$\sigma$. 
The integrated velocity ranges are indicated in the bottom-right corner. 
Black contours represent the distribution of 1.3~mm continuum emission. 
The CO outflow lobe from Figure~\ref{fig:outflow} is also overlaid. 
The systemic velocity of 16.0~$\mathrm{km}~\mathrm{s^{-1}}$ was determined by fitting this \ce{CH3OH} line. 
The filled gray ellipse in the lower-left corner indicates the size of the synthesized beam. 
\label{fig:sw2}}
\end{figure}

The central protobinary system, C1 and C2, is not the only outflow-driving source in IRAS 08303--4303; 
the NE and SW2 condensations also drive molecular outflows. 
Since bipolar outflows are closely associated with accretion disks, we might expect to detect disks in these components as well. 
The NE condensation does not show rich molecular line features, while the SW2 condensation displays several COM lines, although the lines are much weaker than those observed in the central region (see Figure~\ref{fig:spec} and Table~\ref{tab:line}). 
Figure~\ref{fig:sw2} demonstrates the velocity structure of the \ce{CH3OH} line around the SW2 condensation. 
There is a velocity gradient in the northeast--southwest direction, which is roughly parallel to the major axis of SW2 but perpendicular to the associated CO bipolar outflow. 
Therefore, considering that the SW2 condensation is about 1600$\times$500~au in size (Table~\ref{tab:cont_prop}), it seems to be a circumstellar disk around the embedded high-mass (proto)star. 
Since this condensation is marginally resolved in the QUARKS observations, higher angular resolution (and sensitivity) observations are required to investigate its velocity structure in detail. 

%% file: 4_1_kinematic.tex
\begin{figure*}[htbp]
    \centering
    \includegraphics[width=\textwidth]{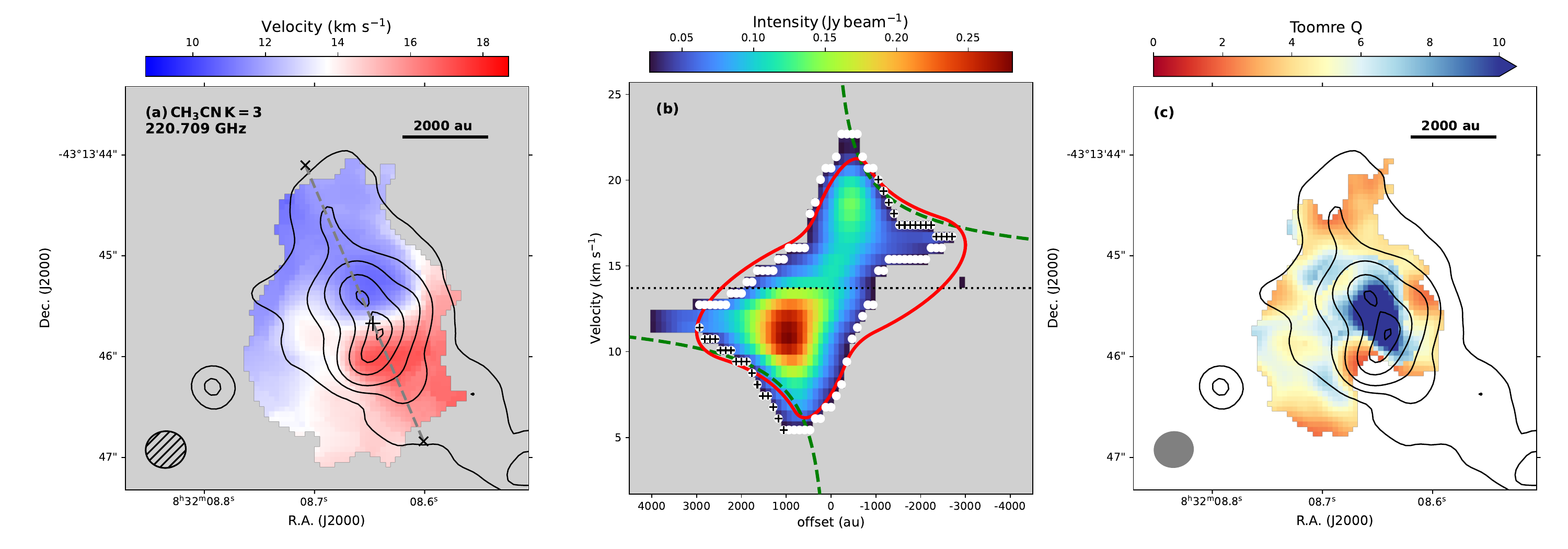}
    \caption{\textbf{(a)} \ce{CH3CN} $J=12\text{--}11~K=3$ line intensity-weighted velocity map. 
    The image is masked below $5\sigma$. 
    Black contours represent the distribution of 1.3~mm continuum emission. 
    The black hatched ellipse in the lower-left corner indicates the size of the synthesized beam. 
    \textbf{(b)} Position--velocity (PV) diagram along the gray dashed line in (a). 
    The image is masked below $7\sigma$. 
    The offset is measured from the black cross in (a), which marks the rotation center. 
    Black crosses mark the data points from the $7\sigma$ emission edge used for fitting a pure Keplerian rotation using Keplerfit. 
    The best-fit velocity model corresponds to a $61~M_\sun$ system and is shown as the green dashed lines. 
    White dots mark the data points from the $7\sigma$ emission edge used for the combined velocity model of Keplerian rotation and free-fall motion. 
    The red solid line shows the best-fit combined velocity model for a $32~M_\sun$ system. 
    In both cases, the inclination angle is fixed at $55\arcdeg$. 
    The dotted horizontal line indicates the systemic velocity, 13.7~$\mathrm{km}~\mathrm{s^{-1}}$. 
    \textbf{(c)} Distribution of the Toomre $Q$ parameter for $M_\mathrm{enc} = 32~M_\sun$, derived from the 1.3~mm continuum map and the temperature distribution in Appendix~\ref{sec:appendixA}, using Equation~\ref{eq:ToomreQ}. 
    \label{fig:diskkin}}
\end{figure*}

The total binary mass (enclosed mass, $M_\mathrm{enc}$) can be derived from the rotation kinematics of the circumbinary disk combined with the inclination information. 
The inclination of the circumbinary disk was determined earlier by fitting the integrated line intensity map, as shown in Figure~\ref{fig:mom0}, yielding 55\arcdeg. 
Assuming Keplerian rotation, the enclosed mass is given by 
\begin{equation}
    M_\mathrm{enc}=\frac{v^2r}{G}\frac{1}{\sin^2 i}, 
\end{equation}
where $v$ is rotation velocity at radius $r$, $G$ the gravitational constant. 
The systemic velocity of the central protobinary system was measured to be $13.7~\mathrm{km~s}^{-1}$ by fitting the 234.187~GHz \ce{SO2} line profile, which shows similar intensities toward the C1 and C2 condensations (see Figure~\ref{fig:mom0}). 
For other molecules exhibiting asymmetric intensity distributions, the fitted peak velocities are considerably closer to the velocity of C1 than to the actual systemic velocity. 
The rotation center is defined as the point with the systemic velocity in the intensity-weighted velocity map along the line connecting the C1 and C2 continuum peaks. 
Using this information, together with the \ce{CH3CN} ($J=12\text{--}11~K=3$) position--velocity (PV) diagram in Figure~\ref{fig:diskkin}, we employed a Keplerian velocity distribution model fitter (Keplerfit)  to derive $M_\mathrm{enc}$ \citep{2019A&A...629A..10B, 2023ascl.soft08012B}. 
A $7\sigma$ emission edge was used for the fitting after excluding the inner region within [–1000, +1000]~au. 
Because the circumbinary disk does not contain actual mass at the rotation center, the rotation at small radius may deviate from Keplerian motion, and this region was therefore masked. 
The outer region beyond 3000~au was also masked because of the limited velocity resolution ($\sim1.3~\mathrm{km~s^{-1}}$), which prevents the velocity structure at low velocities from being resolved. 
The data points used in Keplerfit are marked with black crosses in Figure~\ref{fig:diskkin}(b). 
The inclination, systemic velocity, and rotation center were fixed, while the mass was set as the only free parameter. 
The enclosed mass was estimated to be $61\pm5~M_\sun$ from the best fit, which is delineated by the green dashed lines. 

We also performed an additional fitting of the \ce{CH3CN} ($J=12\text{--}11~K=3$) PV diagram using the method of \citet{2022MNRAS.509..748W}. 
This fitting considers both Keplerian rotation along the tangential direction and free-fall motion in the radial direction. 
The LSR velocity within the disk is then expressed as 
\begin{equation}
\begin{split}
    v_\mathrm{LSR} =\ & v_\mathrm{sys} + \sin i \times \left[ \sqrt{GM_\mathrm{enc}} \frac{x}{R^{3/2}} \right. \\
    & \left. + \sqrt{2GM_\mathrm{enc}} \frac{z}{R^{3/2}}\right], 
    \label{eq:velocity}
\end{split}
\end{equation}
where $x$ and $z$ are coordinates centered at the rotation center. 
Here, $x$ lies in the disk plane, perpendicular to the line of sight, while $z$ also lies in the disk plane but perpendicular to $x$. 
The distance from the rotation center is $R=\sqrt{x^2+z^2}$, which is constrained to the range [$R_\mathrm{in},~R_\mathrm{out}$]. 
We fixed $v_\mathrm{sys} = 13.7~\mathrm{km~s^{-1}}$, $i = 55\arcdeg$, and [$R_\mathrm{in},~R_\mathrm{out}$] = [1000, 3000]~au, consistent with the Keplerfit procedure, and treated the mass as the only free parameter. 
We then fit $v_\mathrm{LSR}$ as a function of $x$ using the white dots in Figure~\ref{fig:diskkin}(b), selected from the $7\sigma$ emission edge within 3000~au. 
We found that a model with an enclosed mass of $32\pm2~M_\sun$ best reproduces the data. 
The fitting result is shown as the red solid line in Figure~\ref{fig:diskkin}(b). 
Having determined that the enclosed mass range is between 32 and $61~M_\sun$ according to the two methods, we can derive the individual component masses by measuring the distance from the rotation center to each condensation peak. 
The C1 peak is located 630~au from the rotation center, while the C2 is 270~au away. 
By definition of the center of mass, the distance ratio from the rotation center is the inverse of the mass ratio. 
Based on the observed distance ratio of 2.3:1, we estimated a mass ratio of 1:2.3. 
This yields a primary star mass in C2 of 22--43~$M_\sun$ and a secondary star mass in C1 of 10--18~$M_\sun$. 

We assessed the gravitational stability of the circumbinary disk using the central stellar mass, 1.3~mm continuum emission, and temperature information, specifically through the Toomre $Q$ parameter \citep{1964ApJ...139.1217T}. 
This parameter evaluates the balance between the local gravitational forces of the disk and the stabilizing effects of shear and thermal pressure \citep{2023A&A...677A.171A}. 
Following \citet{2023A&A...677A.171A}, we write Q as 
\begin{equation}
\begin{split}
    Q =\ & \frac{c_s \Omega}{\pi G \Sigma}, \\
    c_s =\ & \sqrt{\frac{\gamma k_\mathrm{B}T}{\mu m_\mathrm{H}}}, \\
    \Omega =\ & \sqrt{\frac{G(M_\ast+M_\mathrm{disk}(\leq r))}{r^3}}, \\
    \Sigma =\ & \frac{S_{\nu} R_\mathrm{gd}}{B_{\nu}(T_\mathrm{d}) \Omega_B \kappa_{\nu}}. 
    \label{eq:ToomreQ}
\end{split}
\end{equation}
Here, $c_s$ denotes the sound speed within the disk, $\Omega$ is the Keplerian angular velocity, and $\Sigma$ is the surface density of the disk. 
The adiabatic index is $\gamma = 7/5$, $k_\mathrm{B}$ is the Boltzmann constant, $\mu$ is the mean molecular weight of the gas (assumed to be 2.8), and $m_\mathrm{H}$ is the mass of a hydrogen atom. 
$M_\ast$ represents the mass of the central protostar(s), while $M_\mathrm{disk}(\leq r)$ is the enclosed disk mass within radius $r$, which is neglected here because the stellar-to-disk mass ratio is very high ($M_\ast/M_\mathrm{disk}\gtrsim 15$). 
Finally, $S_\nu$ is the continuum intensity, and $\theta_B$ is the beam solid angle. 
If the Toomre $Q$ parameter is smaller than 2, the disk is gravitationally unstable. 
It is also roughly proportional to $M_\ast/M_\mathrm{disk}$, and disks with masses larger than 10–20\% of the stellar mass tend to be gravitationally unstable \citep{2016ARA&A..54..271K, 2023A&A...677A.171A}. 
So, with $M_\mathrm{enc} \sim 32$--61~$M_\sun$ and a relatively small circumbinary disk mass $M_\mathrm{disk} \sim 2~M_\sun$, this circumbinary disk is overall gravitationally stable. 
However, Figure~\ref{fig:diskkin}(c), the distribution of the calculated Toomre $Q$ parameter for $M_\mathrm{enc} = 32~M_\sun$, indicates that certain regions---especially southeast of C2---are gravitationally unstable ($Q < 2$). 
Additionally, given that the disk midplane may be significantly colder than the gas temperature used in the calculation, other regions could also be gravitationally unstable. 
For $M_\mathrm{enc} = 61~M_\sun$, the values would be approximately 1.5 times higher. 

%% file: 4_2_previous.tex
To date, only a few high-mass protobinary systems have been identified. 
IRAS 20126+4104 was the first of these systems, consisting of a primary (proto)star with a mass of $\sim$$7~M_\sun$ and a companion located at a projected separation of around 850~au \citep{2005ApJ...631L..73S}. 
Another candidate high-mass binary system is G11.92--0.61~MM1 with a projected separation of about 2000~au \citep{2018ApJ...869L..24I}. 
The primary component has a mass of roughly $\sim$$34~M_\sun$, while the secondary is a low-mass companion ($<0.6~M_\sun$), likely formed through disk fragmentation. 
G35.20--0.74~N--B contains a binary system with a companion mass of $\sim$$6~M_\sun$, an apparent separation of roughly 800~au from the primary component with a mass $\gtrsim11~M_\sun$, and is surrounded by a circumbinary disk traced by \ce{CH3CN} \citep{2013A&A...552L..10S, 2016A&A...593A..49B}. 
The IRAS 16547--4247--E protobinary system has a total enclosed mass of about $25~M_\sun$ and an apparent separation of $\sim$300~au \citep{2019ApJ...872..176Z}. 
For this system, individual circumstellar disks as well as the circumbinary disk have been identified \citep{2020ApJ...900L...2T}. 
W3(\ce{H2O}) is another example of a high-mass protobinary system where both a circumbinary disk and individual circumstellar disks have been observed \citep{2018A&A...618A..46A}. 
This system consists of two $\sim10~M_{\sun}$ protostars at a projected separation of $\sim$2300~au, surrounded by a circumbinary disk spanning $\sim$6000~au in diameter. 
Notably, the kinematics of its circumbinary disk deviates significantly from typical Keplerian rotation. 
NGC~7538~IRS1 is a high-mass binary system with a projected separation of $\sim$430~au, and its total luminosity corresponds to a stellar mass of about $30~M_\sun$ \mbox{\citep{2017A&A...605A..61B}}. 
IRAS 07299--1651 features a close binary system with component masses of $12.5~M_\sun$ and $10~M_\sun$, with an apparent separation of only $\sim$180~au \citep{2019NatAs...3..517Z}. 
Finally, IRAS 17216--3801 is a well-characterized high-mass binary system with component masses of approximately $20~M_\sun$ and $18~M_\sun$, with a circumbinary disk and a close projected separation of $\sim$170~au, as observed in near-infrared emission \citep{2017ApJ...835L...5K}. 

Among the aforementioned high-mass protobinary systems, IRAS 16547--4247--E is the most extensively studied target regarding its circumbinary disk. 
\citet{2019ApJ...872..176Z} identified a circumbinary disk traced by \ce{CH3OCHO} and determined its enclosed mass through kinematic analysis to be $25\pm 3~M_{\sun}$ under the assumption of Keplerian rotation at an inclination of $55 \pm 5\arcdeg$. 
The radius of this disk is $1190 \pm 290~\mathrm{au}$ at an adopted distance of 2.9~kpc. 
Higher-resolution ($\sim$0\farcs05) observations by \citet{2020ApJ...900L...2T} revealed two individual circumstellar disks in the \ce{H2O} and refractory molecular (e.g., SiS, NaCl, and SiO isotopologues) lines, which are rotating roughly in the opposite directions. 
Besides, the circumbinary disk was observed by traditional disk tracers such as \ce{CH3CN} and \ce{SO2} similarly to IRAS 08303--4303--C in this study (Figure~\ref{fig:mom1}). 
Unfortunately, we cannot detect any circumstellar disk within the circumbinary disk of IRAS 08303--4303--C mainly due to the limited angular resolution of the QUARKS observations. 
The protobinary system of IRAS 08303--4303--C has a larger separation of $\sim$900~au, a larger disk radius of about 3000~au, and a larger enclosed mass of 32--61~$M_\sun$ compared to the IRAS 16547--4247--E. 
Thus, it can be considered a scaled-up version of IRAS 16547--4247--E. 
However, despite its larger enclosed mass, the bolometric luminosity of IRAS 08303--4303 ($\sim$$10^4~L_\sun$) is one order of magnitude smaller, compared to the $\sim$$10^5~L_\sun$ of IRAS 16547--4247. 

If a primary component (C2) has a mass between 22 and $43~M_\sun$, its bolometric luminosity is expected to be similar to or greater than $10^5~L_\sun$ at the zero-age main sequence (ZAMS) \citep{1996ApJ...460..914V, 2005sf2a.conf..351M, 2011MNRAS.416..972D}. 
However, its actual observed luminosity ($\sim$$7000~L_\sun$) is at least an order of magnitude lower. 
One explanation for this discrepancy is that a high accretion rate has caused an increase in the stellar radius while decreasing the surface temperature and luminosity. 
According to \citet{2009ApJ...691..823H} and \citet{2013ApJ...772...61K}, the ``bloating" phase of massive protostars is characterized by a high accretion rate, ranging from a few $10^{-4}$ to a few $10^{-3}~M_{\sun}~\mathrm{yr}^{-1}$, alongside a stellar radius of several $100~R_\sun$. 
From Section~\ref{subsec:result-outflow}, the mass accretion rate calculated from CO outflows exceeds $5\times10^{-4}~M_{\sun}~\mathrm{yr}^{-1}$, and could be even higher if the outflow inclination angle is greater than 57\arcdeg, which is consistent with this ``bloating" phase. 
\citet{2022MNRAS.509..748W} also discussed this phenomenon in relation to the high stellar mass (40--70~$M_\sun$) observed in G19.01--0.03~MM1, which has a moderate luminosity of approximately $10^4~L_\sun$, a situation similar to that presented in this work. 
In addition, the presence of unresolved components could contribute to this discrepancy. 
For instance, according to Table~1 in \citet{2011MNRAS.416..972D}, a single 30~$M_\sun$ star has $L\gtrsim10^5~L_\sun$, whereas two 15~$M_\sun$ stars have $L < 4\times10^4~L_\sun$. 
If the estimated primary protostellar mass in C2 is distributed among two or more unresolved sources, the resulting luminosity would be lower, consistent with the observations. 
In fact, this system is further resolved into multiple individual components in recent higher-resolution observations \citep{2026ApJ..1006...25Y}, which favors this possibility although it is not known how many of the components contain embedded (proto)stars. 

%% file: 4_3_streamer.tex
\begin{figure*}[htbp]
\centering
\includegraphics[width=\textwidth]{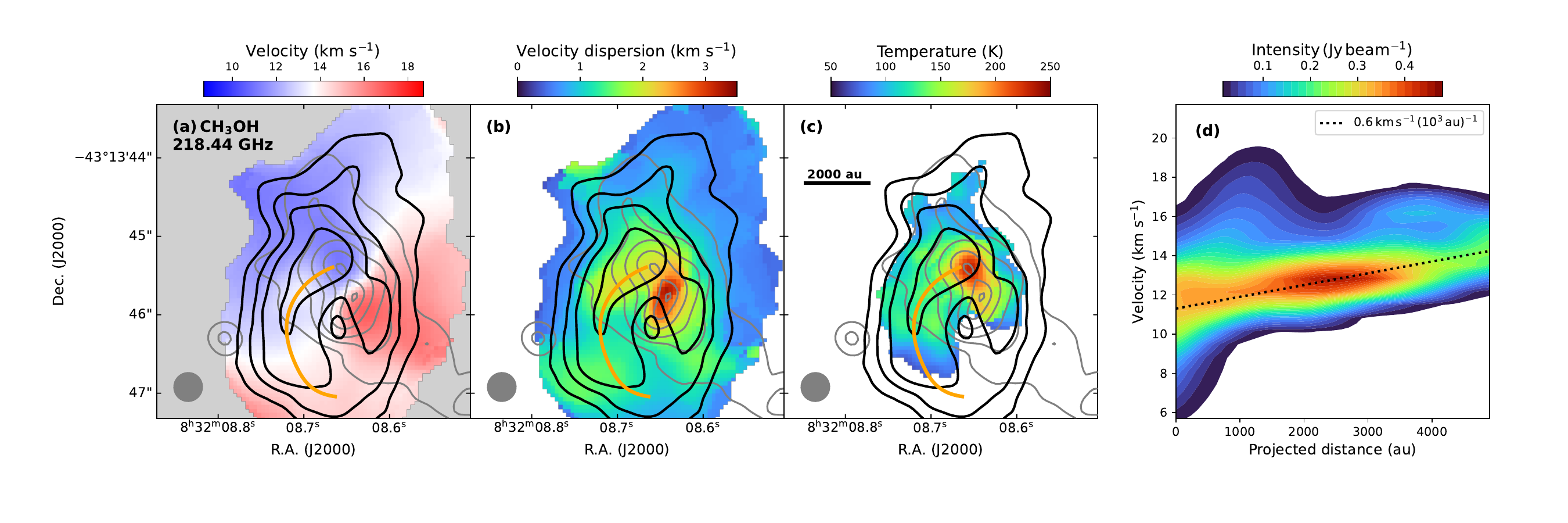}
\caption{\textbf{(a)} \ce{CH3OH} $4_{2,3}\text{--}3_{1,2}~E$ line intensity-weighted velocity map. 
    Gray contours represent the 1.3~mm continuum emission. 
    The image is masked below the $5\sigma$ level. 
    The black contours show the integrated intensity map of the same transition, and the orange line marks the candidate ``streamer,'' obtained by fitting the emission ridge of the integrated intensity map. 
    The gray ellipse in the lower-left corner indicates the synthesized beam size. 
    \textbf{(b)} Same as (a), but with the velocity dispersion map shown as the color image. 
    \textbf{(c)} Same as (a), but with the temperature distribution as the color image, derived from fitting multiple \ce{CH3OH} transitions. 
    For details, see Appendix~\ref{sec:appendixA}. 
    \textbf{(d)} PV diagram along the orange line in (a). 
    The offset is measured from a position near the C1 continuum peak. 
    The red dashed line indicates the velocity gradient estimated by visual inspection. 
\label{fig:streamers}}
\end{figure*}

Recently, nonspherical inflows referred to as streamers have been discovered both in low- and high-mass star-forming regions. 
These streamers transport materials from the inner envelope to the protostellar disk \citep{2022A&A...667A..12V, 2023ApJ...956...82F, 2025A&A...701A.141L}. 
We also detected a streamer candidate around the circumbinary disk. 
Figure~\ref{fig:mom0} illustrates an arc-like structure extending from the C1 condensation, which is observed in various molecules such as \ce{CH3CN}, \ce{HNCO}, \ce{CH3OH}, and \ce{H2CO}. 
The \ce{H2CO} has been utilized to trace streamers around low-mass YSOs \citep{2022A&A...667A..12V}, while \ce{CH3OH} and \ce{CH3CN} have been employed to trace streamers around MYSOs \citep{2023ApJ...956...82F, 2023ApJ...959L..31O, 2025SciA...11.4512O}. 
A velocity gradient of $0.6~\mathrm{km~s^{-1}~(10^3~au)^{-1}}/\sin i$ is observed along the streamer, accompanied by an increase in velocity dispersion toward C1 (Figure~\ref{fig:streamers}). 
A binary system associated with a streamer resembles the low-mass binary system [BHB2007] 11, where spiral-like infalling streams link the circumbinary disk and the envelope \citep{2017A&A...603L...3A}. 
However, such smaller-scale streamers, which connect the circumbinary disk to the individual circumstellar disks in [BHB2007] 11 \citep{2019Sci...366...90A}, are not detected here presumably due to the coarse resolution. 
The low-mass triple system IRAS 04239+2436 also exhibits spiral streamers produced by gravitational interactions among its members \citep{2023ApJ...953...82L}. 

The C1 condensation is weaker than C2 in 1.3~mm continuum emission, 
while it exhibits higher integrated intensities than the latter in most molecular lines, as in Figure~\ref{fig:spec}. 
This difference may be attributed to differences in temperature and/or molecular abundance. 
To investigate this further, we made a rotational temperature map using the intensities of multiple \ce{CH3OH} transitions, employing the assumption of optically thin line emission in LTE (for detailed methodology, see Appendix~\ref{sec:appendixA}). 
The results indicate that the region surrounding C1 has a higher temperature compared to C2 (Figure~\ref{fig:streamers}). 
The temperature of the streamer ranges from 50 to 150~K, increasing as it approaches C1. 
This value is much higher than the 10--15~K typically found in low-mass streamers \citep{2020NatAs...4.1158P, 2025ApJ...985..166K}, and comparable to the $\sim$100~K high-mass streamer reported by \citet{2025SciA...11.4512O}. 
The molecular abundances toward C1 are notably higher than those of C2; 
specifically, the abundances of \ce{CH3CN} and \ce{CH3OH} are 5–7 times greater (see Appendix~\ref{sec:appendixA}). 
One possible explanation for C1’s elevated temperature and molecular richness is the impact of materials accreting via the streamer onto the disk plane, which may lead to accretion shock. 

It would be worthwhile to calculate both the mass of the streamer and the mass accretion rate through it. 
However, the spiral structure resembling the streamer is not clearly visible in the 1.3~mm continuum emission. 
This suggests that the weak continuum emission detected in the streamer region may originate from an envelope along the line of sight, rather than being directly associated with the streamer itself. 
Consequently, calculating the mass from continuum observations is challenging. Similarly, estimating the mass using \ce{CH3OH} or \ce{H2CO}---molecular lines that trace the streamer---is also difficult, as their abundances vary strongly with environmental conditions. 
Another issue is that performing an accurate kinematic analysis is complicated by the unknown inclination of the streamer. 
If the streamer lies in the plane of the disk, we could assume an inclination of 55\arcdeg, but this assumption is problematic, as there have also been reports of streamers located off the disk midplane \citep{2025A&A...701A.141L}. 
Future higher-sensitivity observations would facilitate mass estimation, while higher angular- and velocity-resolution studies would enable a detailed analysis of the streamer and its interface with the disk. 

%% file: 4_4_nature.tex
Binary systems generally form through two primary mechanisms: disk fragmentation and turbulent fragmentation. 
In the former, gravitational instability within a disk leads to the formation of close binaries ($<500$~au), typically resulting in aligned circumstellar disks \citep{1994MNRAS.269L..45B, 2010ApJ...708.1585K, 2019Sci...366...90A}. 
In contrast, wide binaries ($>500$~au) usually form through turbulent fragmentation, which produces systems with misaligned circumstellar disks due to the initially random angular momentum distribution in a turbulent core \citep{2010ApJ...725.1485O, 2017NatAs...1E.172L}. 
For instance, \citet{2018MNRAS.475.5618B} showed in clump-scale numerical simulations that multiple systems formed in such environments often possess misaligned disks, some of which are associated with circumbinary material. 
These disordered systems are consistent with observations of the binary protostar IRS 43, which comprises a circumbinary disk encompassing two misaligned circumstellar disks ($\gtrsim 60\arcdeg$), attributed to either turbulent fragmentation or recent dynamical perturbations \citep{2016ApJ...830L..16B}. 
Despite such initial misalignment, a circumbinary disk can form \textit{a posteriori} through the subsequent accretion of gas from the surrounding envelope. 
\citet{2018MNRAS.475.5618B} also demonstrated that the orientation of a circumbinary disk is governed by the angular momentum of the infalling material, even when it is significantly misaligned with the existing binary systems. 
Circumbinary disks form naturally around binary systems when the specific angular momentum of the infalling material is sufficiently large \citep{1997MNRAS.285...33B, 2019ApJ...871...36M}. 
In such cases, circumbinary disks are established on short timescales, typically within a few orbital periods of the binary system \citep{2005ApJ...623..922O, 2019ApJ...871...36M}. 

The projected separation of $\sim$900~au in IRAS 08303--4303--C indicates that this system is a wide binary. 
The PAs of the C1 and C2 condensations are nearly perpendicular (Table~\ref{tab:cont_prop}). 
While the PA of an individual continuum condensation may not perfectly reflect the true orientation of the circumstellar disk due to potential influence from unresolved substructures, the fact that the outflows from C1 and C2 also appear to be oriented in approximately perpendicular directions further corroborates this misalignment. 
Such misaligned disk--outflow systems suggest that this wide binary formed through turbulent fragmentation on a large scale. 
The circumbinary disk might form subsequently after the binary formation. 
On the other hand, additional unresolved components could have formed within the circumbinary disk due to disk fragmentation. 
Notably, the C2 condensation appears highly elongated toward the southeast from its continuum peak (see, e.g., Figure~\ref{fig:diskkin}(a)). 
This non-Gaussian feature hints the presence of an unresolved component on the southeastern side. 
Moreover, the low Toomre $Q$ parameter in this region (see Figure~\ref{fig:diskkin}(c)), along with the discrepancies in the observed luminosity and the derived enclosed mass discussed in Section~\ref{subsec:discuss-compare}, also suggests the possible existence of an unresolved component. 

As noted above, \citet{2026ApJ..1006...25Y} resolved the C1 and C2 condensations into at least three distinct components each, one of which lies southeast of the C2 peak. 
While the limited resolution of our QUARKS data precluded detailed analysis of individual disk kinematics, they marginally detected a circumstellar disk candidate toward one component within C1 in \ce{H2O} line emission, which often serves as a more reliable tracer for hot, inner disk dynamics than COMs \citep{2017NatAs...1E.146H, 2018ApJ...860..119G, 2020ApJ...900L...2T}. 
However, their molecular line analysis focused solely on the single component. 
Therefore, a joint analysis of our comprehensive data with higher-resolution data, including theirs, is needed to characterize disk rotation in the other components and accretion flows onto individual circumstellar disks. 
Although the continuum condensation and outflow morphologies observed in this study appear to support the misalignment among the individual components, unresolved substructures might produce some uncertainty in the measured orientations of the innermost disks and their associated outflows. 
Such a joint analysis would help measure their true orientations and test our interpretation of the misalignment and turbulent fragmentation. 

%% file: 5_Summary.tex
We carried out ALMA observations of the high-mass star-forming region IRAS 08303--4303 in Band 6 with an angular resolution of about 0\farcs3, as part of the QUARKS project. 
We detected multiple dense condensations and the associated disk-outflow systems, and investigated the physical and dynamical properties of them. 
The main results are summarized as follows. 

\begin{enumerate}
\item We detected six dense continuum condensations at 1.3~mm. 
At least four of these condensations (NE, C1, C2, SW2) show signs of star formation activity, as indicated by the line-rich spectra and associated outflows. 
The NE, C2, and SW2 condensations demonstrate clear bipolar molecular outflows, while the C1 (and possibly SW1) condensation is associated with weak outflow features. 

\item The C1 and C2 condensations, located apparently $\sim$900~au apart, form a wide protobinary system and reside within a single rotating structure. 
This structure is characterized as a `circumbinary disk' and exhibits a velocity pattern similar to Keplerian rotation, with higher velocities observed near the center. 
Multiple molecules, including \ce{CH3CN}, \ce{SO2}, \ce{CH3OH}, \ce{HC3N}, \ce{HNCO}, and \ce{OCS}, trace this rotation. 

\item The total mass of the protobinary system was derived using both Keplerian rotation and combined Keplerian rotation–infall models. 
The total mass estimates range between 32 and 61~$M_\sun$, according to two different fitting results. 
The masses of the primary and secondary protostars were computed to be 22--43 and 10--18~$M_\sun$, respectively, based on their distances from the rotation center. 

\item The temperature of the circumbinary disk was determined from a rotational diagram analysis using multiple transitions of \ce{CH3OH}, yielding a value of approximately 130~K. 
The mass of the circumbinary disk was then estimated to be about 2~$M_\sun$ based on the continuum flux and temperature data. 
Given the relatively low disk mass compared to the high protostellar mass, the disk is overall gravitationally stable, except for the southeastern region of the C2 condensation showing $Q<2$. 

\item From the outflow parameters, the mass accretion rate onto the primary protostar in C2 was derived to be greater than $5\times10^{-4}~M_\sun~\mathrm{yr^{-1}}$. 
This high accretion rate may explain the discrepancy between the observed bolometric luminosity of less than $10^4~L_\sun$ and the relatively high estimated protostellar mass of 22--43~$M_\sun$. 
During the protostellar stage, a high accretion rate can cause the stellar surface to expand, resulting in a lower luminosity compared to the zero-age main sequence. 
In addition, the possible presence of an unresolved component around C2 could also account for this inconsistency. 

\item The C1 condensation is brighter in most molecular lines and exhibits a higher temperature compared to the primary component, C2. 
This may be attributed to a streamer connected to C1, which is observed in several molecular lines, including \ce{CH3OH}, \ce{H2CO}, \ce{CH3CN}, and \ce{HNCO}. 
Accreting material flowing along the streamer may interact with the disk plane, leading to enhanced temperatures and molecular abundances around C1 due to accretion shock. 
\end{enumerate}

The IRAS 08303--4303 region provides valuable insights into the rarely observed circumbinary disks around massive YSOs. 
It serves as an excellent laboratory for future studies aimed at understanding high-mass multiple formation through higher-resolution observations. 

%% file: 6_acknowledgments.tex
We thank the anonymous referee for his/her constructive and thoughtful comments.
This paper makes use of the following ALMA data: ADS/JAO. ALMA\#2019.1.00685.S and 2021.1.00095.S. 
ALMA is a partnership of ESO (representing its member states), NSF (USA) and NINS (Japan), together with NRC (Canada), MOST and ASIAA (Taiwan, China), and KASI(Republic of Korea), in cooperation with the Republic of Chile. 
The Joint ALMA Observatory is operated by ESO, AUI/NRAO, and NAOJ. 
GG and LB gratefully acknowledge support by the ANID BASAL project FB210003. 
HLL is supported by the National Natural Science Foundation of China (No.\,12673033), by Yunnan Fundamental Research Project (grant No.\,202301AT070118, 202401AS070121), and by Xingdian Talent Support Plan-Youth Project. 
DL is supported by the National Natural Science Foundation of China (NSFC) Programs No.\,12588202. 
SRD acknowledges support from the Fondecyt Postdoctoral fellowship (project code 3220162) and ANID BASAL project FB210003. 
Suinan Zhang acknowledges support from the Strategic Priority Research Program of the Chinese Academy of Sciences (CAS) Grant No.\,XDB0800303 and the NAOJ ALMA Scientific Research Grant Code 2025-29B. 
MJ acknowledges the support of the Research Council of Finland Grant No.\,348342. 

%% file: 7_Temprot.tex
\begin{figure*}[htbp]
    \centering
    \includegraphics[width=\textwidth]{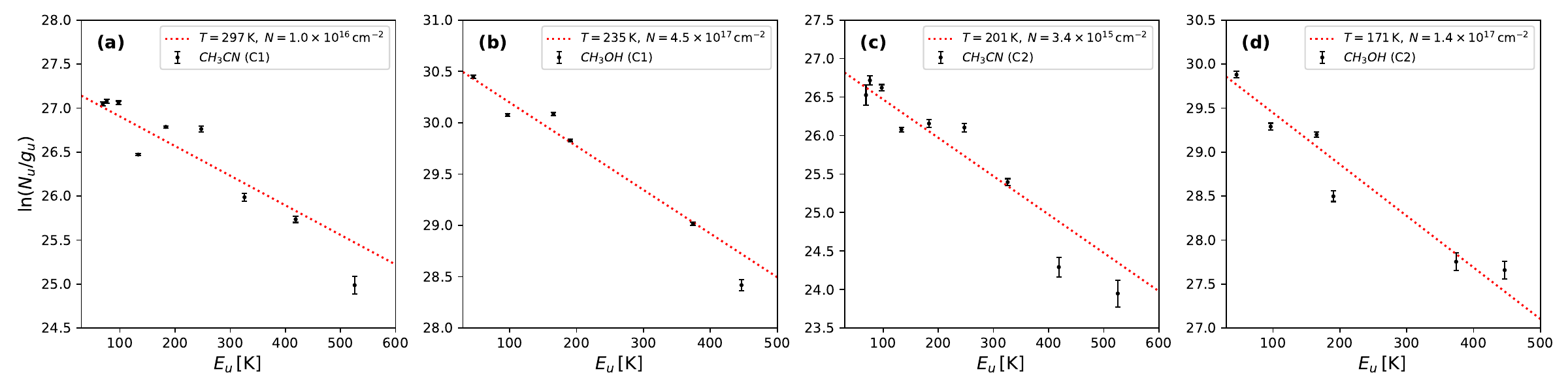}
    \caption{
    \textbf{(a)} Rotational diagram analysis of condensation C1 using \ce{CH3CN}, assuming optically thin emission. 
    \textbf{(b)} Same as (a), but for \ce{CH3OH}. 
    \textbf{(c)}--\textbf{(d)} Same as (a)--(b), but for condensation C2. 
    \label{fig:rotD}}
\end{figure*}

Assuming the LTE condition, the rotational temperature ($T_\mathrm{rot}$) can be calculated using the equations provided by \cite{1999ApJ...517..209G}, as shown below: 
\begin{equation}
\begin{split}
    & N_u^{\mathrm{thin}} = \frac{8 \pi k_\mathrm{B} \nu^2 W}{h c^3 A_{ul}}, \\
    & \ln{\frac{N_u^{\mathrm{thin}}}{g_u}} + \ln{C_{\tau}} = \ln{N} - \ln{Z} - \frac{E_u}{k_\mathrm{B} T_\mathrm{rot}}. 
    \label{eq:Trot}
\end{split}
\end{equation}
Here $\nu$ is the rest frequency; 
$W$ the integrated line intensity; 
$h$ the Planck constant; 
$c$ the speed of light; 
$A_{ul}$ the Einstein coefficient; 
$N_u^{\mathrm{thin}}$ the upper-level column density when the lines are optically thin; 
$g_u$ the upper-level degeneracy; 
$C_{\tau} = \tau/(1-e^{-\tau})$ the optical depth correction factor ($C_{\tau} \simeq 1$ when $\tau \ll 1$); 
$N$ the total column density of the molecule; 
$Z$ the partition function; 
and $E_u$ the upper-level energy. 
From Table~\ref{tab:line}, nine transitions of \ce{CH3CN} and six of \ce{CH3OH} were detected within the QUARKS frequency band, each corresponding to a different upper-level energy. 
By plotting $\ln(N_u^{\mathrm{thin}}/g_u)$ against $E_u / k_\mathrm{B}$, the inverse of the slope yields the rotation temperature, while the y-intercept allows us to determine the total column densities of \ce{CH3CN} and \ce{CH3OH} (see Figure~\ref{fig:rotD}). 
For \ce{CH3OH}, we performed the rotation diagram analysis by combining the transitions of A-type and E-type species. 
Similar methodologies have been adopted in several recent studies where the number of detected \ce{CH3OH} lines was insufficient to determine the $A/E$ ratio independently \citep{2020MNRAS.498L..87B, 2022ApJ...939...84B, 2023A&A...678A..18F}. 
This simplification is further justified by the high-temperature environment ($T > 37$~K) where the $A/E$ ratio is expected to approach unity within 0.5\% \citep{2011A&A...533A..24W}. 
The hot-core nature of our target ($T > 100$~K) significantly exceeds the $A/E$ equilibration threshold, supporting the single excitation temperature assumption for both species. 
Molecular spectroscopic data (e.g., rest frequencies, Einstein coefficients, upper-level degeneracies, and partition functions) were obtained from the CDMS catalog \citep{2001A&A...370L..49M, 2005JMoSt.742..215M, 2016JMoSp.327...95E}. 
Partition functions were interpolated to the appropriate temperatures. 

\begin{deluxetable*}{lcccccc}[htbp]
\tablecaption{Results from the rotational diagram analysis: 
rotational temperature ($T_\mathrm{rot}$); 
molecular line column density ($N_\mathrm{line}$); 
\ce{H2} column density ($N_\mathrm{H_2}$); 
molecular abundance ($X_\mathrm{line}$); 
and optical depth ($\tau$) of representative transitions. 
\label{tab:temp}}
\tablehead{ & \multicolumn{6}{c}{\textbf{\ce{CH3CN}}}\\
    \colhead{Condensation} & \colhead{$T_\mathrm{rot}$} & \colhead{$N_\mathrm{CH_3CN}$} & \colhead{$N_\mathrm{H_2}$ } & \colhead{$X_\mathrm{CH_3CN}$} & 
    \colhead{$\tau_{J=12\text{--}11~K=3}$} & \colhead{$\tau_\mathrm{J=12\text{--}11~K=5}$}\\
    & (K) & ($10^{15}~\mathrm{cm^{-2}}$) & ($10^{23}~\mathrm{cm^{-2}}$) & ($10^{-8}$) & & }
\startdata
\textbf{C1} & $297\pm8$ & $10.0\pm0.6$ & $2.7\pm1.0$ & $3.8\pm1.4$ & 1.43 & 0.44 \\
\textbf{C2} & $201\pm7$ & $3.4\pm0.3$ & $5.2\pm1.9$ & $0.6\pm0.2$ & 0.74 & 0.19 \\
\textbf{SW2} &  &  &  &  &  &  \\
\textbf{Central} & $175\pm6$\tablenotemark{a} & $1.26\pm0.09$\tablenotemark{a} & $1.1\pm0.4$ & $1.2\pm0.4$ & 0.33 & 0.08 \\
\hline
 & \multicolumn{6}{c}{\textbf{\ce{CH3OH}}}\\
& $T_\mathrm{rot}$ & $N_\mathrm{CH_3OH}$ & $N_\mathrm{H_2}$ & $X_\mathrm{CH_3OH}$ 
    & $\tau_{4_{2,3}\text{--}3_{1,2}}$ & $\tau_{10_{2,9}\text{--}9_{3,6}}$ \\
    & (K) & ($10^{16}~\mathrm{cm^{-2}}$) & ($10^{23}~\mathrm{cm^{-2}}$) & ($10^{-6}$) & & \\
\hline
\textbf{C1} & $235\pm3$ & $45\pm1$ & $3.4\pm1.2$ & $1.4\pm0.5$ & 0.79 & 0.38 \\
\textbf{C2} & $171\pm6$ & $14\pm1$ & $6.1\pm2.2$ & $0.22\pm0.08$ & 0.42 & 0.18 \\
\textbf{SW2} & $262\pm9$ & $9.1\pm0.6$ & $1.3\pm0.5$ & $0.7\pm0.3$ & 0.13 & 0.07 \\
\textbf{Central} & $127\pm2$ & $8.9\pm0.3$ & $1.5\pm0.6$ & $0.6\pm0.2$ & 0.48 & 0.16 \\
\hline
\enddata
\tablenotetext{a}{The \ce{CH3CN} $J=12\text{--}11~K=8$ transition was not used due to its weak intensity.}
\end{deluxetable*}

\begin{figure*}[htbp]
    \centering
    \includegraphics[width=\textwidth]{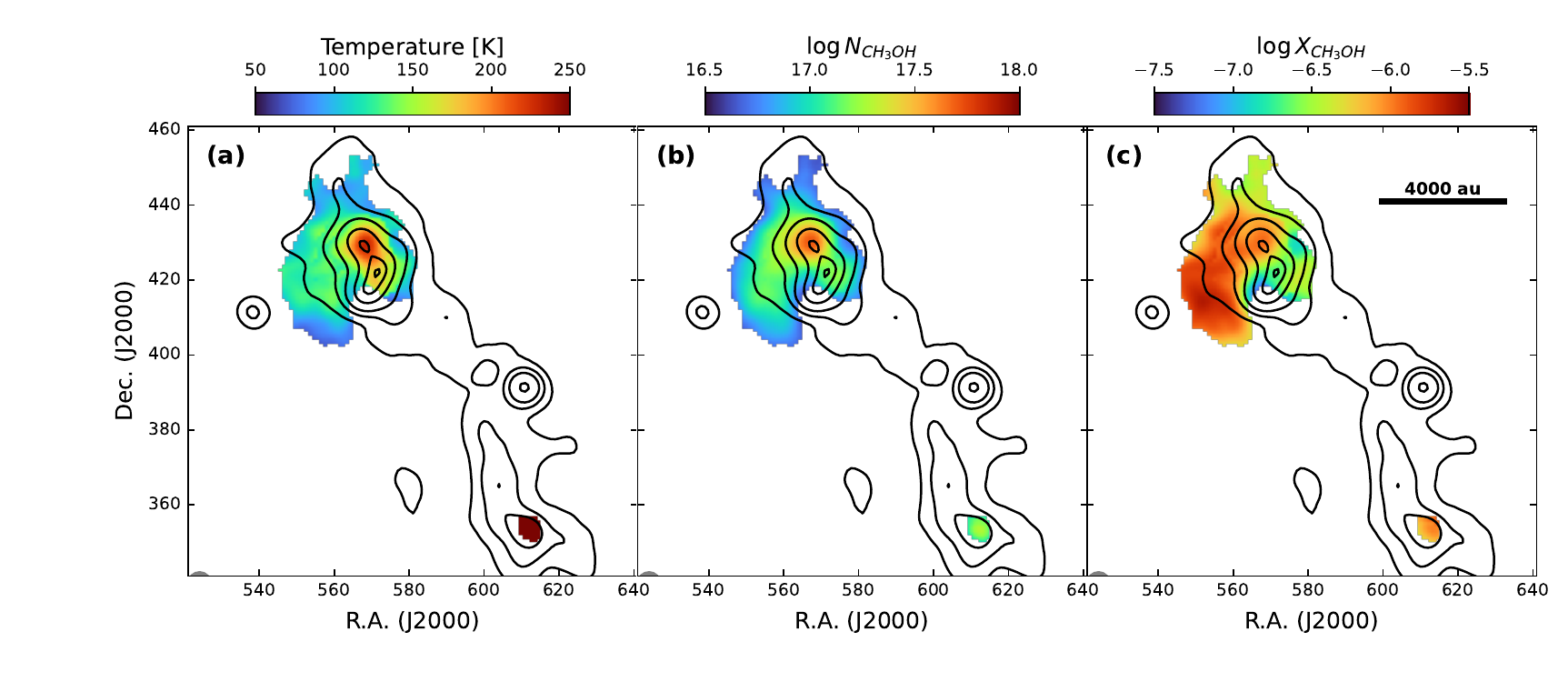}
    \caption{
    \textbf{(a)} Temperature distribution derived from \ce{CH3OH} rotational diagram analysis. 
    \textbf{(b)} Column density distribution of \ce{CH3OH}. 
    \textbf{(c)} Abundance distribution of \ce{CH3OH} relative to \ce{H2}. In each panel, contours delineate the distribution of 1.3~mm continuum emission.
    \label{fig:rotmap}}
\end{figure*}

All the rotational temperature and molecular column density calculations for C1, C2, SW2, and the overall central protobinary system were performed based on the optically thin assumption, where $C_{\tau}$ is considered to be unity for all lines. 
The \ce{CH3CN} $J=12\text{--}11~K=0$ and $K=1$ transitions are partly blended due to their close rest frequencies, and the \ce{CH3CN} $J=12\text{--}11~K=6$ transition overlaps with nearby \ce{CH3^{13}CN} and HNCO lines. 
Therefore, Gaussian fitting was used to separate the integrated line intensities of these transitions. 
For the SW2 condensation, the calculations were only conducted using \ce{CH3OH} lines because of the weak intensities of the \ce{CH3CN} lines. 
Assuming that the dust temperature ($T_\mathrm{d}$) is equal to the derived rotational temperature, the average \ce{H2} column density within the source solid angle $\theta$ was estimated from the 1.3~mm continuum flux density $F_{\nu}$ using the following equation from \citet{2008A&A...487..993K}: 
\begin{equation}
N_\mathrm{H_2} = \frac{F_{\nu} R_\mathrm{gd}}{B_{\nu}(T_\mathrm{d}) \theta \kappa_{\nu} \mu m_\mathrm{H}}. 
\end{equation}
Molecular abundance was calculated as the ratio of the molecular column density to the \ce{H2} column density. 
From the temperatures and molecular column densities derived from the rotation diagram analysis, we also estimated the optical depths of several \ce{CH3OH} and \ce{CH3CN} transitions using the non-LTE radiative transfer code RADEX \citep{2007A&A...468..627V}. 
The \ce{CH3OH} line emission is more optically thin than that of \ce{CH3CN} toward all regions; 
therefore, we present the temperature and mass estimates derived from \ce{CH3OH} in Table~\ref{tab:cont_prop}. 
All results from the rotation diagram analysis are summarized in Table~\ref{tab:temp}. 
Figure~\ref{fig:rotmap} shows the maps of temperature, molecular column density, and molecular abundance obtained by applying equation~(\ref{eq:Trot}) to each pixel of the integrated line intensity maps of \ce{CH3OH}, where at least four \ce{CH3OH} transitions were detected above $5\sigma$. 

%% file: 8_ch3cn.tex
\begin{figure*}[htbp]
    \centering
    \includegraphics[width=\textwidth]{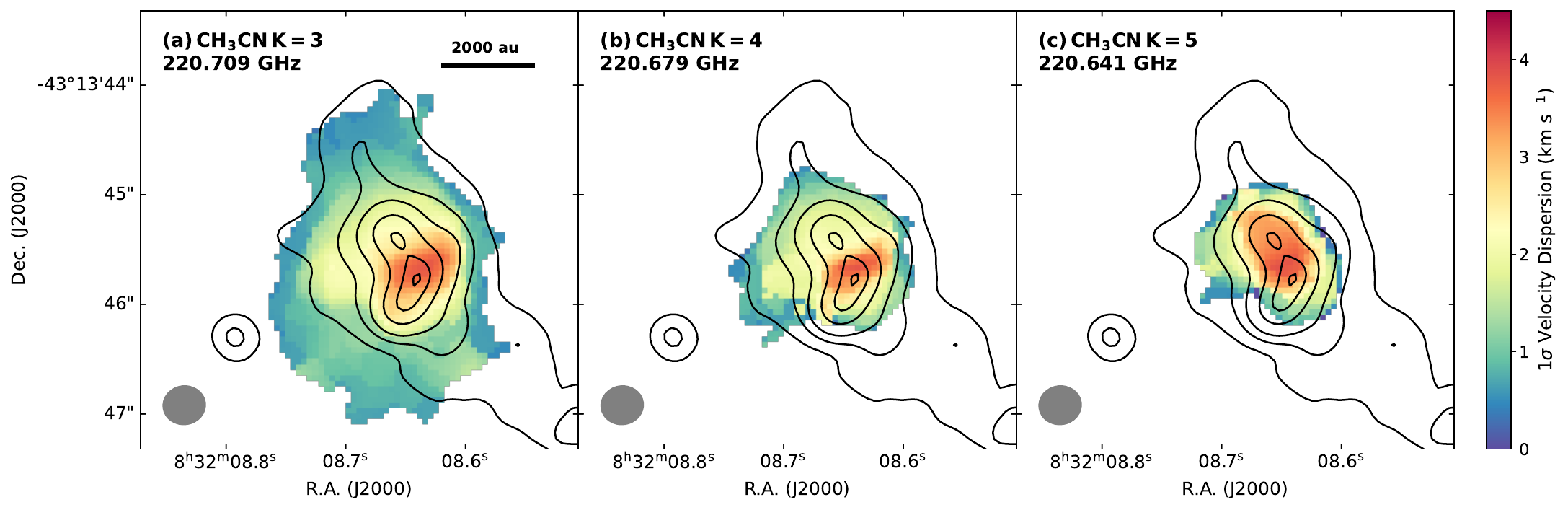}
    \caption{Velocity dispersion ($1\sigma$; Moment 2) maps of the \ce{CH3CN} $K=3$, 4, and 5 transitions, shown in color. 
    Emission below the $5\sigma$ detection threshold is masked in all maps. 
    Black contours show the 1.3~mm continuum emission. 
    The filled gray ellipse in the lower-left corner of each panel indicates the synthesized beam. 
    \label{fig:mom2}}
\end{figure*}

Figure~\ref{fig:mom2} displays the velocity dispersion maps (Moment 2) of 3 \ce{CH3CN} transitions. 
Around the C2 condensation, all three transitions exhibit broad linewidths corresponding to $\mathrm{FWHM} \gtrsim 7~\mathrm{km~s}^{-1}$, assuming Gaussian line profiles. 
Unresolved circumstellar disk rotation may contribute to these broad linewidths. 
Toward the C1 condensation, the $K=5$ transition shows a clearly larger velocity dispersion than the $K=3$ and 4 transitions, which may indicate that it traces a hotter, faster-rotating region closer to the central source. 
Additionally, the $K=3$ and 4 transitions exhibit more extended emission and reveal a localized enhancement in velocity dispersion immediately west of the C1/C2 center, which may be associated with the outflow. 
This possible outflow-related enhancement is not apparent in the more compact emission traced by the $K=5$ transition. 

